# Agentic Inference of Adsorbate Configurations on Catalyst Surfaces from Vibrational Spectra

*Dong Hyeon Mok*[1], *Seungjun Cha*[2,3], *Jaekyum Kim*[4], *Seoin Back*[4,5,6,7,*], *Victor Fung*[3,*] *and Guoxiang Hu*[2,8,*]

[1]Department of Chemical and Biomolecular Engineering, Institute of Emergent Materials, Sogang University, Seoul 04107, Republic of Korea

[2]School of Materials Science and Engineering, Georgia Institute of Technology, Atlanta, GA 30332, USA

[3]School of Computational Science and Engineering, Georgia Institute of Technology, Atlanta, GA 30332, USA.

[4]KU-KIST Graduate School of Converging Science and Technology, Korea University, Seoul 02841, Republic of Korea

[5]Department of Integrative Energy Engineering, Korea University, Seoul 02841, Republic of Korea

[6]Institute for Multiscale Matter and Systems (IMMS), Ewha Womans University, Seoul 03760, Republic of Korea

[7]Center for Hydrogen and Fuel Cells, Korea Institute of Science and Technology (KIST), Seoul 02792, Republic of Korea

[8]School of Chemistry and Biochemistry, Georgia Institute of Technology, Atlanta, GA 30332, USA.

AUTHOR INFORMATION

**Corresponding Author**

E-mail: sback@korea.ac.kr, victorfung@gatech.edu, emma.hu@mse.gatech.edu

**Abstract**

Identifying adsorbate configurations on catalyst surfaces from vibrational spectra is essential for understanding catalytic reaction mechanisms, yet remains challenging because multiple species and binding environments can produce overlapping spectral features. Here, we introduce Spectroscopy-Informed Reasoning for Electrocatalytic Networks (SIREN), a large language model-based agent framework that iteratively infers adsorbate configurations through structural proposals, vibrational calculations and feedback-guided refinement. SIREN employs a machine learning force field to calculate vibrational fingerprints of proposed adsorbate configurations and uses spectral discrepancies to guide subsequent modifications to adsorbate identities, binding sites and spatial arrangements. Across computational benchmarks on four catalyst surfaces, SIREN achieved higher spectral recovery rates than Bayesian optimization alone. Applications to experimental high-resolution electron energy loss spectra of Rh (111) and potential-dependent surface-enhanced Raman spectra of oxide-derived Cu further demonstrated its ability to generate chemically interpretable surface state hypotheses. For $CO_2$ electroreduction, the inferred intermediates were compared with reported product Faradaic efficiencies to explore possible reaction pathways. These results establish SIREN as a framework for translating vibrational spectral evidence into explicit atomistic hypotheses and supporting mechanistic interpretation of catalytic reactions.

## 1. Introduction

Identifying the species present on a catalyst surface under reaction conditions is central to understanding how a reactant is converted into different products. Vibrational spectroscopy provides a way to follow these species as the reaction environment changes[1-3]. For example, Zhan et al. combined in situ surface-enhanced Raman spectroscopy (SERS) with density functional theory (DFT) calculations to identify adsorbed intermediates associated with ethylene and ethanol formation during $CO_2$ electroreduction on Cu[4]. Interpreting such spectra, however, becomes difficult when several adsorbed species coexist. Different adsorbates can contribute bands in the same frequency range, and even a single species can exhibit distinct frequencies as its adsorption environment changes[5]. On Cu (111), increasing CO coverage can populate bridge sites and displace neighboring CO molecules from atop positions, giving rise to a distinct infrared signature[6]. A band that is poorly reproduced by a proposed structure may therefore call for a change in the adsorbate identity, its binding site or the arrangement of its neighbors.

Recovering this structural information from an experimentally measured spectrum is an inverse problem. One approach is to learn the relationship between spectra and selected features of the surface. Lansford and Vlachos used simulated infrared spectra of CO on Pt to train neural networks that inferred distributions of CO binding sites and local coordination environments from experimental spectra[7]. A second approach is to propose explicit structures and refine them against the measurement. Slagle et al. developed ActiveStructOpt, which uses active learning and a graph neural network surrogate to refine atomic structures against X-ray pair distribution functions and absorption spectra[8]. For a reacting surface, an additional challenge arises when both the composition and the arrangement of the adsorbates are unknown. Even on a specified catalyst facet, each plausible mixture admits several site occupations, and a change that improves agreement in one spectral region may compromise a band already reproduced elsewhere.

Choosing which structural change to test next therefore calls for chemical judgment. A researcher draws on knowledge of surface chemistry and the results of previous observations to decide which candidate configuration merits further investigation. Large language model (LLM)-based agents offer a way to incorporate this judgment into an iterative search[9-11]. In our previous work on single-atom catalysts, agents used the calculated effects of earlier active-site modifications to guide subsequent design choices. Through this iterative process, they identified catalysts in which hydrogen bonding selectively stabilized an oxygen reduction intermediate, thereby breaking the conventional adsorption-energy scaling relations[12]. These results demonstrated that LLM-based agents could use computational evidence to make chemically informed decisions during catalyst design[13]. Here, we investigate whether this capability extends to surface spectroscopy, where the task is to propose and refine adsorbate configurations that are consistent with a measured vibrational spectrum.

In this work, we present Spectroscopy-Informed Reasoning for Electrocatalytic Networks (SIREN), an LLM-based framework for inferring candidate adsorbate configurations from a target vibrational spectrum and a structural model of the catalyst surface. The agents propose which species occupy the surface, how many are present and where they bind. A machine learning force field (MLFF) efficiently predicts harmonic vibrational fingerprints, which the agents compare with the target spectrum to refine subsequent proposals. Across computational benchmarks on four catalyst surfaces, SIREN recovered the target fingerprints more frequently than Bayesian optimization alone. We then examined two experimental cases. A high-resolution electron energy loss spectroscopy (HREELS) spectrum on Rh (111) provided a well-characterized reference against which to assess both the proposed adsorbates and their ranking. For potential-dependent SERS spectra on oxide-derived Cu (OD-Cu), we examined how the inferred intermediates changed with applied potential and used the reported product Faradaic efficiencies to assess which reaction pathways were consistent with those changes. Taken together, these results show how SIREN translates vibrational spectra into explicit hypotheses about the identity, coverage, and arrangement of adsorbates on a catalyst surface. When considered with product measurements, these candidate configurations provide a basis for proposing how the observed surface chemistry may be related to competing reaction pathways.

## 2. Results

### 2.1 Overall framework

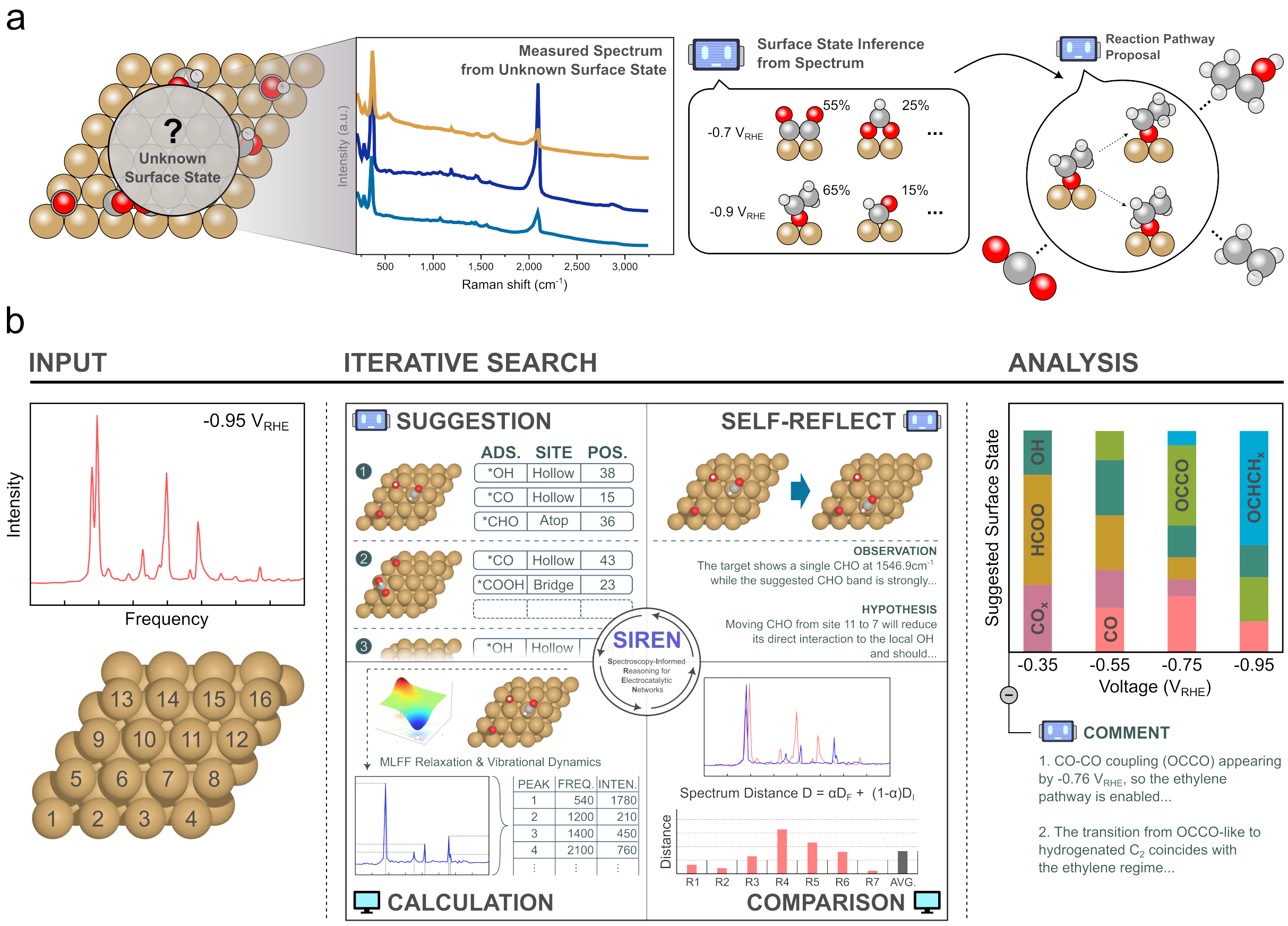

**Figure 1.** Overview of the SIREN framework for spectroscopy-guided inference of adsorbate configurations. (a) Schematic illustration of the inverse problem, in which a vibrational spectrum measured on a catalyst surface is used to propose plausible adsorbate configurations, which may subsequently inform hypotheses about catalytic reaction pathways. (b) Iterative workflow comprising the input spectrum and surface model, candidate proposal, structural relaxation and vibrational calculations using a MLFF, spectral comparison and feedback-guided refinement. Selected adsorbate configurations can be combined with additional reaction information for optional post-hoc mechanistic interpretation.

SIREN is an iterative framework for inferring adsorbate configurations from vibrational spectra by combining language-model reasoning with atomistic simulations (**Figure 1**)[14]. The framework takes a target spectrum and a structural model of the catalyst surface on which the spectrum is assumed to have been acquired. Here, a surface state is defined by the identities, numbers, binding sites and spatial arrangement of adsorbates on the supplied surface. Starting from these inputs, the agents propose candidate surface states, calculate their vibrational fingerprints and compare the results with the target. The resulting discrepancies are incorporated into the search history and used to refine subsequent proposals, while the best-matching candidate identified so far retained as the current best configuration. This process allows the agents to revise both the adsorbate composition and the local arrangement of species as evidence accumulates.

The search combines exploration of alternative adsorbate compositions with refinement of promising configurations. A routing component selects among exploration, exploitation and Bayesian optimization (BO) in the computational benchmarks[15, 16], while a reflection component interprets the calculated results and informs subsequent decisions. For example, a missing spectral feature may motivate the introduction of another adsorbate, whereas a frequency mismatch at an otherwise appropriate composition may motivate a change in binding site or neighboring species. The agents therefore use previous calculations as contextual evidence for deciding which structural changes to test next.

The framework returns candidate surface states consistent with the available spectral evidence. When the input spectrum is acquired during a catalytic reaction, these states represent possible adsorbed intermediates under the corresponding experimental conditions. Additional reaction information can then be used to examine how the inferred intermediates relate to product formation. We evaluated this workflow using computational benchmarks with known target configurations, an experimental HREELS spectrum of co-adsorbates on Rh (111)[17, 18] and potential-dependent SERS spectra acquired during $CO_2$ reduction on OD-Cu[4, 19]. For the latter application, the inferred surface states were subsequently considered together with reported product Faradaic efficiencies (FEs) to develop mechanistic interpretations and compare them with those proposed in the original experimental study.

## 2.2 Computational Benchmarks and Spectral Recovery

We first evaluated SIREN using computationally generated target spectra, for which the underlying adsorbate configurations were known. This controlled benchmark complements the experimental case studies, where the diversity of available reference spectra is limited and the corresponding surface states may themselves be uncertain. For each task, a target configuration was generated randomly and its vibrational spectrum was calculated. The spectrum and the underlying catalyst surface model were then provided to SIREN, while the identities, numbers and positions of the adsorbates in the target surface state were withheld. The task was to identify a configuration that reproduced the target fingerprint within a fixed number of iterations.

To make repeated evaluations practical, we used the UMA MLFF for the atomistic calculations[20, 21]. The applicability of UMA to adsorbate–surface systems has been demonstrated in its original study through benchmarks of energy and force prediction on the Open Catalyst 2020 (OC20) dataset and adsorption configuration optimization using the AdsorbML benchmark. Prior to vibrational analysis, adsorbate positions were optimized using UMA-predicted forces while keeping the surface atoms fixed (**SI Note B.2**). Harmonic vibrational analysis was performed using the Vibrations implementation in the Atomic Simulation Environment (ASE)[22], with the surface atoms held fixed during the vibrational calculations. These calculations provided vibrational modes and frequencies, and the resulting spectra were used as computational vibrational fingerprints.

We assessed the reliability of the MLFF-calculated frequencies against DFT calculations using single-adsorbate configurations spanning 19 adsorbate species and 50 randomly generated multi-adsorbate configurations on Cu (111). The calculated band positions generally followed the DFT results, with the most pronounced discrepancies appearing in the high-frequency region above 3000 $cm^{-1}$, particularly for O-H stretching modes (**Figure S3** and **Table S5**). Within the evaluated structures, these deviations preserved the broad separation between the principal vibrational regions, although they could affect the discrimination of closely spaced features. The observed agreement supports the use of UMA for computationally efficient harmonic vibrational fingerprint calculations for the investigated configurations.

The inference benchmark covered Cu (111), $Cu_3Pd$ (111), $Fe_3Cu$ (111) and Cu (211). Target states were generated from four representative adsorbates relevant to $CO_2$ reduction (CO, COOH, CHO and OH )[23, 24]. Their numbers, combinations and surface positions were varied randomly, and ten target tasks were evaluated for each surface. Each search was allowed up to 50 iterations. Candidate spectra were compared with the target using a distance measure that combined the difference in frequency distributions with differences in integrated spectral weight across predefined frequency regions (**SI Note D.1**)[25]. Successful spectral recovery was defined as a spectral distance of 0.01 or less. This operational threshold was supported by an analysis of sampled configuration pairs, for which the smallest observed distance between non-equivalent states was 0.0108 (**SI Note D.2**)

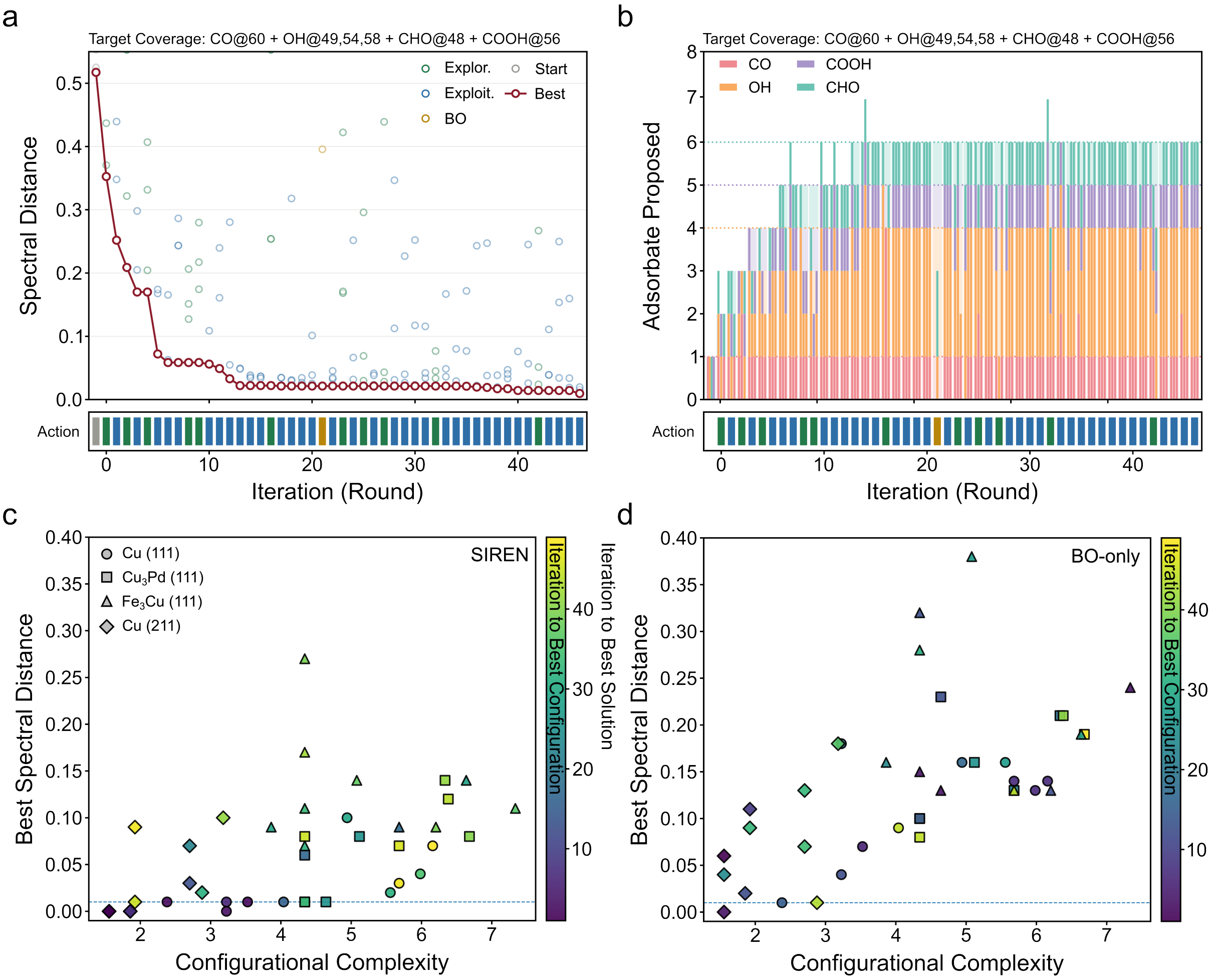

**Figure 2.** Performance of SIREN in computational benchmarks. (a) Representative run showing the spectral distance from the target. Point colors distinguish exploration, exploitation and BO proposals. The strip below the plot indicates the action selected in each iteration. (b) Evolution of the adsorbate composition proposed during the same search. (c, d) Best spectral distance for individual targets versus configurational complexity, obtained with (c) SIREN and (d) BO-only. Symbol shapes distinguish the four catalyst surfaces, and symbol colors denote the iteration at which the best configuration was found. The dashed horizontal line indicates the recovery criterion (D = 0.01). Results for additional language models are provided in **Figure S7**.

**Table 1.** Performance of SIREN and BO-only in the computational benchmarks on four surfaces (ten targets per surface). The average best spectral distance is the mean of the minimum distances reached in individual runs. The success rate is the percentage of runs reaching $D \leq 0.01$ within 50 iterations. Results for additional LLMs are given in **Table S15**.

| | System | Average Best Spectral Distance | Success Rate |
|---|---|---|---|
| SIREN (GPT-5.6-luna) | Cu (111) | 0.030 | 50% |
| | $Cu_3Pd$ (111) | 0.066 | 30% |
| | $Fe_3Cu$ (111) | 0.128 | 0% |
| | Cu (211) | 0.032 | 50% |
| | All | **0.064** | **32.5%** |
| BO-only | Cu (111) | 0.112 | 10% |
| | $Cu_3Pd$ (111) | 0.151 | 0% |
| | $Fe_3Cu$ (111) | 0.211 | 0% |
| | Cu (211) | 0.080 | 10% |
| | All | 0.139 | 5% |

Because the targets differed in both composition and configurational freedom, we characterized configurational complexity using $\log_{10}(N_{config})$, where $N_{config}$ denotes the number of possible adsorbate configurations under the adopted counting scheme. BO was applied to the same target states using the same maximum number of iterations[14, 26]. Across the benchmarks, SIREN reached lower final spectral distances and recovered target fingerprints more frequently than BO alone (**Figure 2** and **Table 1**). Lower-complexity targets were typically solved with smaller spectral distances and in fewer iterations. Nevertheless, complexity alone did not determine difficulty. Some Cu (211) targets remained challenging despite relatively low configurational complexity, which reflects the fewer adsorption sites in the surface model. This behavior is consistent with the importance of local adsorption environments on a stepped surface, where differences in site coordination can affect the spectral response.

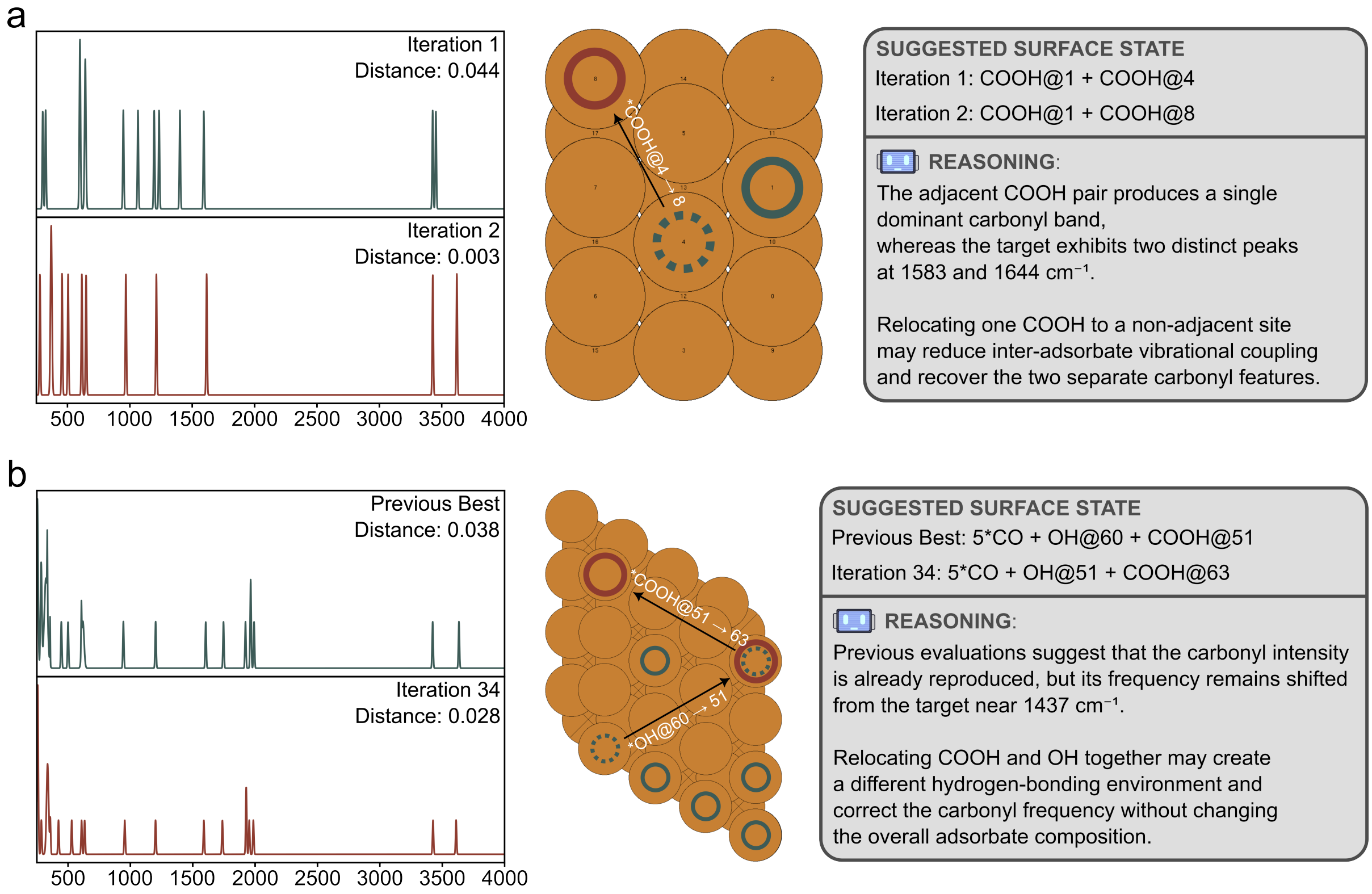

**Figure 3.** Representative feedback-guided revisions during SIREN inference. (a) Relocating one COOH adsorbate while retaining the two COOH composition changes the local adsorbate environment. (b) A joint relocation of OH and COOH targets a residual mismatch near 1437 $cm^{-1}$ while retaining the overall composition.

Individual trajectories illustrate how chemical reasoning was used to guide structural refinement. In **Figure 3a**, a candidate containing two COOH adsorbates failed to reproduce two distinct target features near 1583 and 1644 $cm^{-1}$. The agent proposed moving one COOH from site 4 to site 8 while retaining the other at site 1, hypothesizing that changing the local interaction between the adsorbates would improve the separation of the carbonyl features. The resulting spectral distance decreased from 0.044 to 0.003. Thus, the target fingerprint was recovered by changing the spatial arrangement without changing the adsorbate composition.

**Figure 3b** shows a refinement informed by earlier evaluations. At iteration 34, the agent retained a composition of five CO, one OH and one COOH but jointly relocated OH and COOH. Its interpretation was that the remaining mismatch near 1437 $cm^{-1}$ arose from the local environment of the carbonyl-containing adsorbate, while other features were already reproduced satisfactorily. The proposed rearrangement reduced the distance from 0.038 to 0.028. This example demonstrates how the search history and regional spectral discrepancies can guide a targeted modification late in the search.

These examples show that SIREN can translate spectral discrepancies into specific changes in adsorbate arrangement and evaluate those changes computationally. They also suggest that chemical prior knowledge and accumulated feedback plausibly contribute to the observed search performance. Additional examples of the agents' reasoning are provided in the **Figure S8**.

## 2.3 Case study: HREELS on Rh (111)

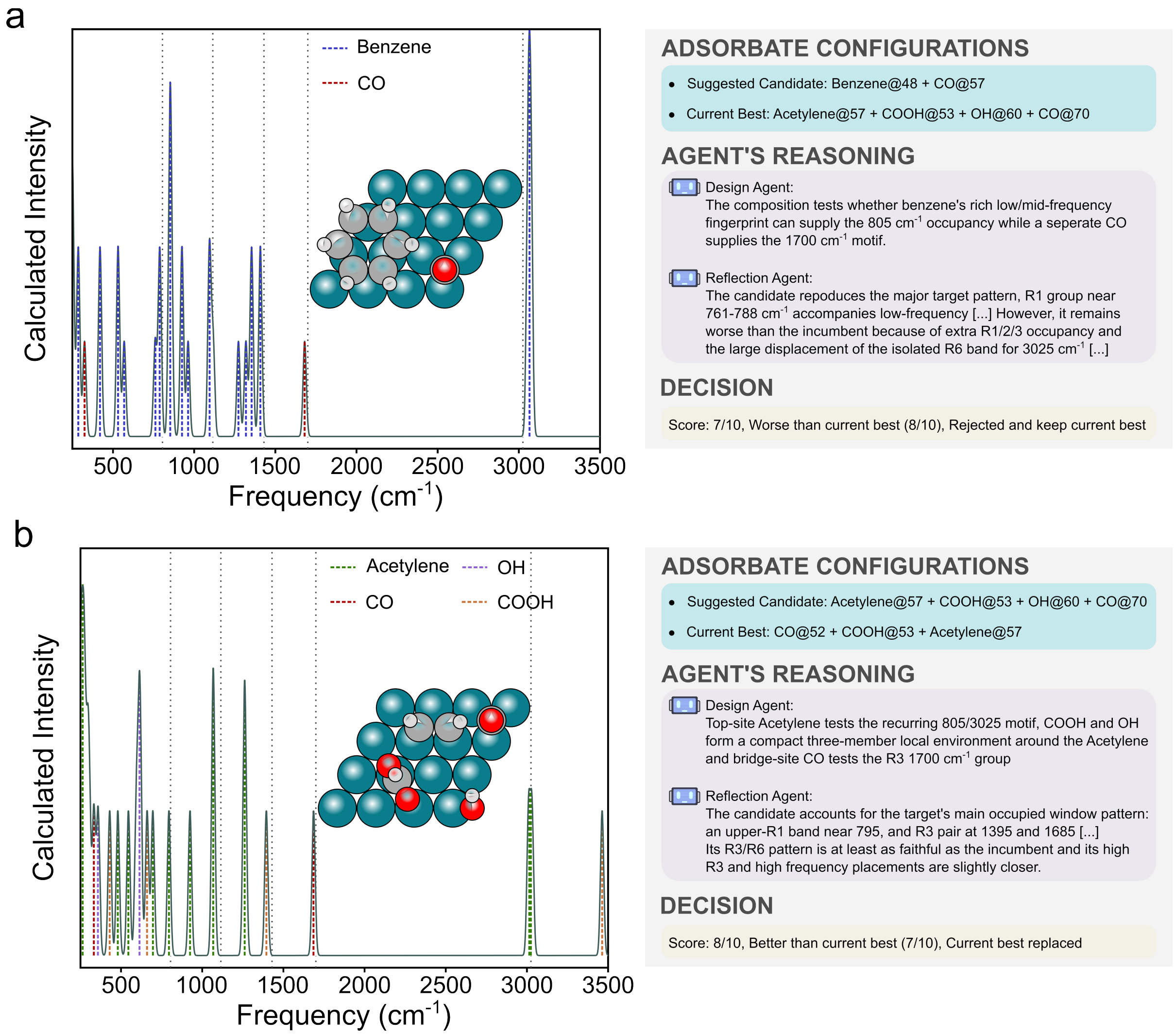

**Figure 4.** SIREN inference from an experimental HREELS spectrum of Rh (111). (a) Representative candidate containing benzene and CO, which reproduced the adsorbate identities of the reported reference state but was not selected as the best-matching configuration. (b) Representative higher-rated alternative containing acetylene and additional co-adsorbates. Gray vertical dashed lines indicate peak positions in the experimental HREELS reference. The experimentally measured spectrum and the reported reference structure are shown in **Figure S5**.

As an experimental case study with a well-characterized reference structure, we used the HREELS spectrum reported by Mate and Somorjai for an ordered Rh (111) (3 × 3) – ($C_6H_6$ + 2CO) co-adsorbate structure at 300 K[17, 18]. The experimental spectrum was supplied as an image together with a Rh (111) surface model and basic information about the measurement. Although the reported reference structure was based on a (3 × 3) unit cell, a (4 × 4) Rh (111) surface was used for inference to mitigate spatial constraints on adsorbate placement. The calculated vibrational fingerprints for comparable configurations showed nearly identical peak positions and ordering between the two cell sizes. Information identifying the reference adsorbate composition was excluded from the input. The candidate library included molecular adsorbates discussed in the reference study, so the search was not restricted to benzene and CO. The candidate species, experimental inputs and search settings are detailed in **SI Note F.1** and **Table S12**, with the experimental reference spectrum shown in **Figure S5**. For this application, spectral agreement was assessed by the reflection agent through comparisons between the experimental image and calculated fingerprints[27]. Unlike in the computational benchmarks, the spectral distance (D) was not used here, because the calculated spectral weights and the experimental intensities were not directly comparable. The agent considered the presence and arrangement of spectral features and their relative prominence when comparing candidates with the current best configuration.

Across the five runs, we compared the candidates that most closely resembled the reference state with those that received the highest agent ratings (**Figure 4**). The first group comprises candidates that carried the adsorbate identities of the reference surface state, namely one benzene and one CO on the (4 × 4) cell. The agent associated the feature near 1700 $cm^{-1}$ with adsorbed CO and interpreted features near 805 and 3025 $cm^{-1}$ as evidence for benzene ring vibrations and aromatic C-H stretching, respectively. These candidates reproduced the two adsorbate identities in the reference surface state, although their benzene-to-CO ratio of 1:1 Was lower than the 1:2 ratio in the reference structure. However, they received coarse ratings of 4–7 out of 10 and were judged inferior to the current best configuration at the time (**Figure 4a**). The second group comprises the candidates the agent rated highest. Nine candidates across the five runs reached the top rating of 8/10. Only two of them contained benzene, whereas six contained acetylene, with the ~1700 $cm^{-1}$ band supplied by CO or by the C=O stretch of CHO (**Figure 4b**). The preference for alternative configurations may partly arise from ambiguities in the calculated vibrational fingerprints, particularly the overlapping frequency assignments of benzene, acetylene, CO and CHO. In the UMA harmonic spectra, the measured 3025 $cm^{-1}$ band is matched more closely by the ≡C–H stretch of acetylene (median 3012 $cm^{-1}$ over all candidates from the five runs) than by the aromatic C–H stretch of benzene (median 3066 $cm^{-1}$), and the ~1700 $cm^{-1}$ band is reproduced equally well by CO (median 1669 $cm^{-1}$ on top sites and 1697 $cm^{-1}$ on bridge sites) and by the carbonyl stretch of CHO (median 1680 $cm^{-1}$). The discrepancy between compositional recovery and the agent's ranking highlights the limitations of matching calculated vibrational fingerprints to experimental HREELS data. Differences in mode-dependent scattering intensities, line shapes and the experimental background or noise can influence the apparent agreement.

The HREELS case therefore demonstrates both a capability and a limitation of the framework. SIREN generated and correctly rationalized a benzene-CO candidate from the spectral evidence in most runs, yet its selection procedure, which can only compare harmonic band patterns with a measured spectral envelope, did not identify the reference surface state as the preferred solution and instead favored acetylene- or CHO-based alternatives in four of the five runs. Improving the correspondence between calculated and measured responses, together with the reliability of candidate ranking, will therefore be important for extending successful fingerprint matching to accurate experimental surface state inference[27, 28].

### 2.4 Case study: SERS on Oxide-Derived Cu and Reaction Pathway Inference

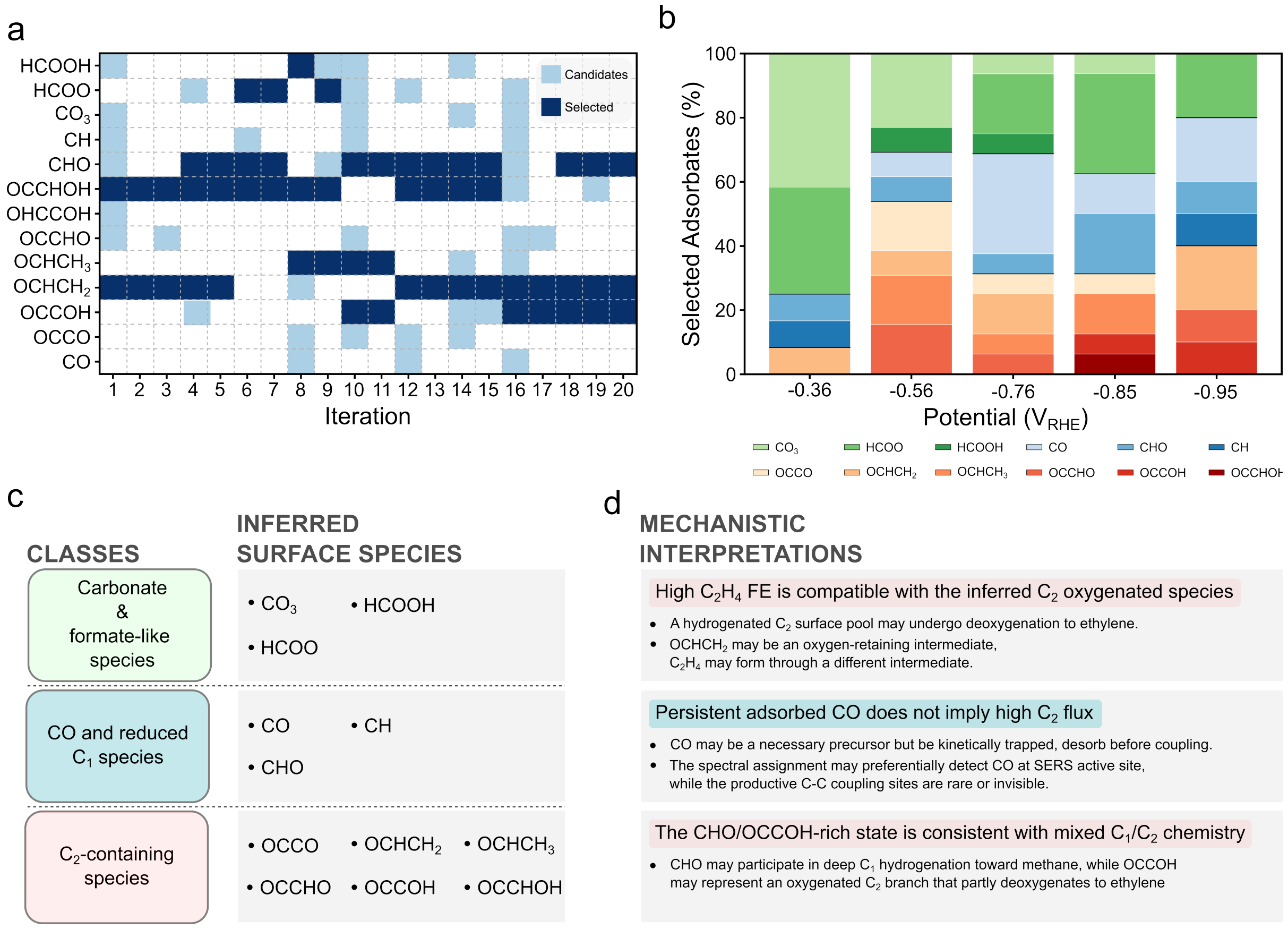

**Figure 5.** Potential-dependent surface state inference and mechanistic interpretation of $CO_2$ electroreduction on OD-Cu. (a) Evolution of adsorbate identities in proposed and selected configurations during a representative inference run at −0.95 $V_{RHE}$. Light blue denotes proposed candidates and dark blue denotes the selected surface state. (b) Fractions of inferred adsorbates at each potential, obtained by pooling adsorbate counts from the final selected configurations of five independent runs. Results obtained with other LLMs are provided in **Figure S9**. (c) Inferred adsorbates grouped into chemical classes. (d) Representative post-hoc mechanistic hypotheses based on the inferred states and their qualitative comparison with published product FEs.

We next applied SIREN to the in situ SERS spectra reported by Zhan et al. for an electrochemically treated Cu electrode during $CO_2$ reduction[4, 19, 29, 30]. Unlike the ordered Rh (111) reference, the surface states responsible for these spectra were not uniquely established by the experimental data. We therefore examined the consistency of the inferred species with the published assignments and used the inferred states as a basis for mechanistic interpretation. The calculations employed an OD-Cu surface model obtained from the reference study and an adsorbate candidate library containing intermediates relevant to $CO_2$ reduction. Further details are provided in the **SI Note F.1**.

Inference was performed for spectra acquired at −0.36, −0.56, −0.76, −0.85 and −0.95 V versus the reversible hydrogen electrode ($V_{RHE}$), with five independent runs at each potential. Within each iteration, three candidate configurations were proposed and assessed relative to the current best configuration. A representative run shows that several alternative adsorbate combinations were explored before the selected composition stabilized after iteration 17 (**Figure 5a**).

To summarize the results, we pooled the adsorbate counts from the final selected surface states of the five runs at each potential and expressed the count of each species as a percentage of the total number of adsorbates in those states (**Figure 5b**). We note that these fractions describe the composition of the inferred configurations, rather than directly measured surface coverages or probabilities of species occurrence. This representation nevertheless allows the recurring chemical character of the inferred surface states to be compared across potentials.

At the least cathodic potential, −0.36 V, the inferred states were dominated by carbonate and formate-like species[19, 31]. At more cathodic potentials, CO and $C_2$-containing candidates became more prominent, along with oxygenated $C_1$ species[3, 32]. The distribution at −0.76 V included a substantial CO contribution, while oxygenated $C_2$ species, including $OCHCH_2$ and OCCOH, contributed to the states inferred at −0.95 V. These changes indicate a broad progression from carbonate and formate-like candidates toward mixtures containing CO and $C_2$ intermediates. Grouping the results into carbonate, formate-like species, CO and reduced $C_1$ species, and $C_2$-containing species makes the broader chemical trend clearer (**Figure 5c**).

The potential-dependent distributions were broadly compatible with the interpretation in the reference studies that CO adsorption and C-C coupling contribute to the development of $C_2$ chemistry, with hydrogenated oxygen-containing intermediates becoming relevant under more cathodic conditions[4, 33-35]. We note that the agreement at this level does not establish a unique atomistic assignment for every band (**Figure S10**). Instead, it indicates that the framework can organize the experimental evidence into candidate surface states whose chemical character can be compared with established mechanistic interpretations.

We subsequently supplied the inferred states together with the product FEs reported in the reference study[4] to an agent tasked with interpreting their relationship (**Figure 5d**). This step constitutes a post-hoc analysis of the inferred surface states rather than part of the inference itself. Where the available FE measurements did not coincide with the spectral acquisition potentials, the comparison was made qualitatively with reference to the neighboring potential range.

The inferred states were broadly consistent with the emergence of $C_2$ chemistry at more cathodic potentials, but their mechanistic interpretation was not identical to that proposed by Zhan et al. SIREN frequently selected candidate configurations containing oxygenated $C_2$ species, including $OCHCH_2$ and $OCHCH_3$, at potentials where substantial ethylene production was also observed. The agent's interpretation suggested that these intermediates might undergo further deoxygenation toward ethylene or participate in a parallel oxygen-retaining pathway. However, the reference study specifically associates OCCO(H)-derived intermediates with ethylene formation and identifies $OCHCH_2$ as a key intermediate in the ethanol-selective pathway[4, 34]. Thus, the coexistence of inferred oxygenated $C_2$ species and ethylene production does not establish their direct participation in ethylene formation. The proposed deoxygenation route remains a hypothesis requiring further validation, whereas the coexistence of distinct $C_2$ reaction pathways is consistent with the reference mechanism.

The mechanistic analysis also highlighted additional relationships between the inferred surface intermediates and product formation. First, CO was repeatedly identified across the investigated potentials, suggesting that its surface presence and its conversion into $C_2$ products may not follow the same potential dependence. The agent therefore proposed that the availability of reactive sites and the kinetics of subsequent C-C coupling steps could influence the observed product distribution. Second, the inference of CHO and OCCOH suggested the possible coexistence of $C_1$ hydrogenation and oxygenated $C_2$ formation pathways[24, 35, 36]. In this interpretation, CHO may participate in further $C_1$ hydrogenation, whereas OCCOH may undergo subsequent reduction along a $C_2$ reaction pathway. These hypotheses provide possible connections between the inferred intermediates and the formation of different reaction products. The full LLM-generated interpretations are provided in **Table S17**.

Taken together, the OD-Cu case study illustrates how potential-dependent surface state inference can be extended to the interpretation of catalytic reaction pathways. By comparing the inferred intermediates with experimentally reported product distributions, the analysis identified both mechanistic interpretations consistent with the reference study and alternative hypotheses concerning oxygenated $C_2$ intermediates, persistent CO and parallel $C_1$/$C_2$ reaction pathways.

**3. Discussion**

Evaluating spectrum-based surface state inference presents an inherent challenge. Computational benchmarks provide known reference configurations, allowing the accuracy of inferred states to be quantitatively assessed. However, because both target and candidate spectra are generated using the same computational model, successful reconstruction primarily demonstrates the framework's ability to solve a self-consistent inverse problem rather than its accuracy in interpreting experimental measurements. Conversely, experimental spectra represent more realistic applications, but the surface configurations are generally unknown and may not be uniquely determined from the available spectral information[7, 8]. Consequently, the accuracy of experimentally inferred surface states cannot be evaluated using the same criteria as the computational benchmarks. These two evaluation approaches therefore provide complementary evidence of the framework's capabilities and limitations.

A further limitation arises from the differences between calculated vibrational fingerprints and experimentally measured spectra. The harmonic approximation and the MLFF employed in this study provide computationally efficient estimates of vibrational frequencies but do not fully reproduce the technique-dependent spectral responses measured by HREELS or SERS[28, 37]. This limitation was particularly evident in the Rh (111) case study[27], where SIREN generated a candidate containing benzene and CO, consistent with the adsorbate identities of the reference structure, but rated alternative acetylene- or CHO-containing configurations more highly. Thus, even when the relevant adsorbates are included among the proposed candidates, discrepancies between calculated and experimental spectra can affect their relative ranking. Similar uncertainties also apply to the OD-Cu application, where differences between the modeled and actual reaction environments may contribute to discrepancies between the proposed mechanistic interpretations and those reported in the reference study. Furthermore, identifying adsorbed intermediates does not directly establish their kinetic contributions to individual reaction pathways.

Despite these limitations, SIREN provides a framework for systematically examining plausible surface states against experimental spectral evidence. Rather than assigning a unique surface state directly from individual spectral features, the framework iteratively proposes and evaluates candidate structures, allowing chemical hypotheses to be revised based on calculated vibrational responses. The inferred configurations can subsequently serve as a basis for examining possible reaction pathways in conjunction with additional experimental observations. Further improvements in the physical accuracy of forward spectral calculations, including a more realistic treatment of the experimental environment and spectroscopic responses, would strengthen the reliability of candidate selection and mechanistic interpretation[23]. In this context, SIREN may serve as a computational aid for interpreting complex surface spectra and formulating testable hypotheses, rather than as a standalone method for definitive surface structure determination.

## 4. Conclusion

In this work, we developed SIREN, a large language model-based agent framework for inferring adsorbate configurations on catalyst surfaces from vibrational spectra through iterative structure proposal, spectral calculation and feedback-guided refinement. Computational benchmarks demonstrated that SIREN could recover target spectral fingerprints across diverse catalyst surfaces and generally achieved better inference performance than Bayesian optimization alone. Representative results further illustrated how the agents used spectral discrepancies and previous evaluations to refine adsorbate compositions and spatial arrangements. Applications to experimental HREELS and SERS spectra demonstrated the feasibility of extending this approach to measured spectral data. For electrochemical $CO_2$ reduction on OD-Cu, the inferred surface states were additionally used to formulate potential-dependent mechanistic hypotheses and to compare them with reported product distributions. Although further improvements in spectral modeling and experimental validation are required, these results establish SIREN as a framework for connecting vibrational spectral evidence with atomistic surface state hypotheses and possible catalytic reaction pathways.

## Code Availability

The code developed in this study is available at (https://github.com/ahrehd0506/SIREN)

## Supporting Information

Additional details on surface and adsorbate models, vibrational calculations and MLFF validation, agent architecture and inference procedures, spectral distance metrics, computational benchmarks, and experimental applications.

## Acknowledgements

D.H.M. acknowledges the support from Korea Institute for Advancement of Technology (KIAT) grant funded by the Ministry of Trade, Industry & Energy (MOTIE), Korea Government (RS-2024-00436106, Human Resource Development Program for Industrial Innovation). S.B. acknowledges the support from the National Research Foundation of Korea (NRF) grants funded by the Korea government (MSIT and MOE) (RS-2024-00448287, RS-2025-16063688, RS-2025-00513832, and RS-2025-02214715), and the generous supercomputing time provided by the Korea Institute of Science and Technology Information (KISTI). G.H. acknowledges support from the U.S. National Science Foundation under Grant # CBET-2442223. This research used resources of the National Energy Research Scientific Computing Center, a DOE Office of Science User Facility supported by the Office of Science of the U.S. Department of Energy under Contract No. DE-AC02-05CH11231 using NERSC award BES-ERCAP0036338.

# Supplementary Information

## Supplementary Note A. Representation of surfaces, adsorbates and spectra

### A.1 Slab Table and Adsorption Site Catalogue

The surface was represented by a plain-text table containing the atom index, element, and nearest-neighbor connectivity of each top-layer atom. Neighbor lists were obtained using Voronoi tessellation (pymatgen VoronoiNN, tolerance 0.5)[1] and converted into an undirected graph for adsorbate adjacency analysis, arrangement signatures, and BO encoding. The slab models are listed in **Table S1** and **Figure S1**.

Only atop sites were considered in the computational benchmarks. For experimental cases, atop sites were identified within 0.8 Å of the highest surface atom; neighboring surface atoms were defined using a cutoff of 1.15 times the shortest surface-atom separation ($d_{nn}$). Bridge sites were the midpoints of neighboring atom pairs, and hollow sites were the centroids of mutually neighboring triples with centroid-to-vertex distances within 0.15 $d_{nn}$ of $d_{nn}/\sqrt{3}$. Hollow sites were classified as hcp when a second-layer atom lay within 0.20$d_{nn}$ laterally beneath the site, and as fcc otherwise. Non-atop sites were assigned indices above the slab atom count. Surface-atom and adsorption-site tables, including site connectivity and mutually excluded sites, were provided to the agents (**Table S4**).

**Table S1. Structural characteristics of the catalyst slab models used in the computational benchmarks and experimental case studies.**

| System | Composition (Atoms) | In-plane Cell | Layers |
|---|---|---|---|
| Cu (111) | $Cu_{64}$ | 4 × 4, a = 10.38 Å | 4 |
| Cu (211) | $Cu_{36}$ | 6.26 × 7.67 Å | 4 |
| $Cu_3Pd$ (111) | $Cu_{48}Pd_{16}$ | 4 × 4, a = 10.52 Å | 4 |
| $Fe_3Cu$ (111) | $Cu_{16}Fe_{48}$ | 4 × 4, a = 10.21 Å | 4 |
| Rh (111) | $Rh_{64}$ | 4 × 4, a = 10.90 Å | 4 |
| OD-Cu | $Cu_{444}O_{48}$ | a = 22.25 Å, γ = 120° | reconstructed |

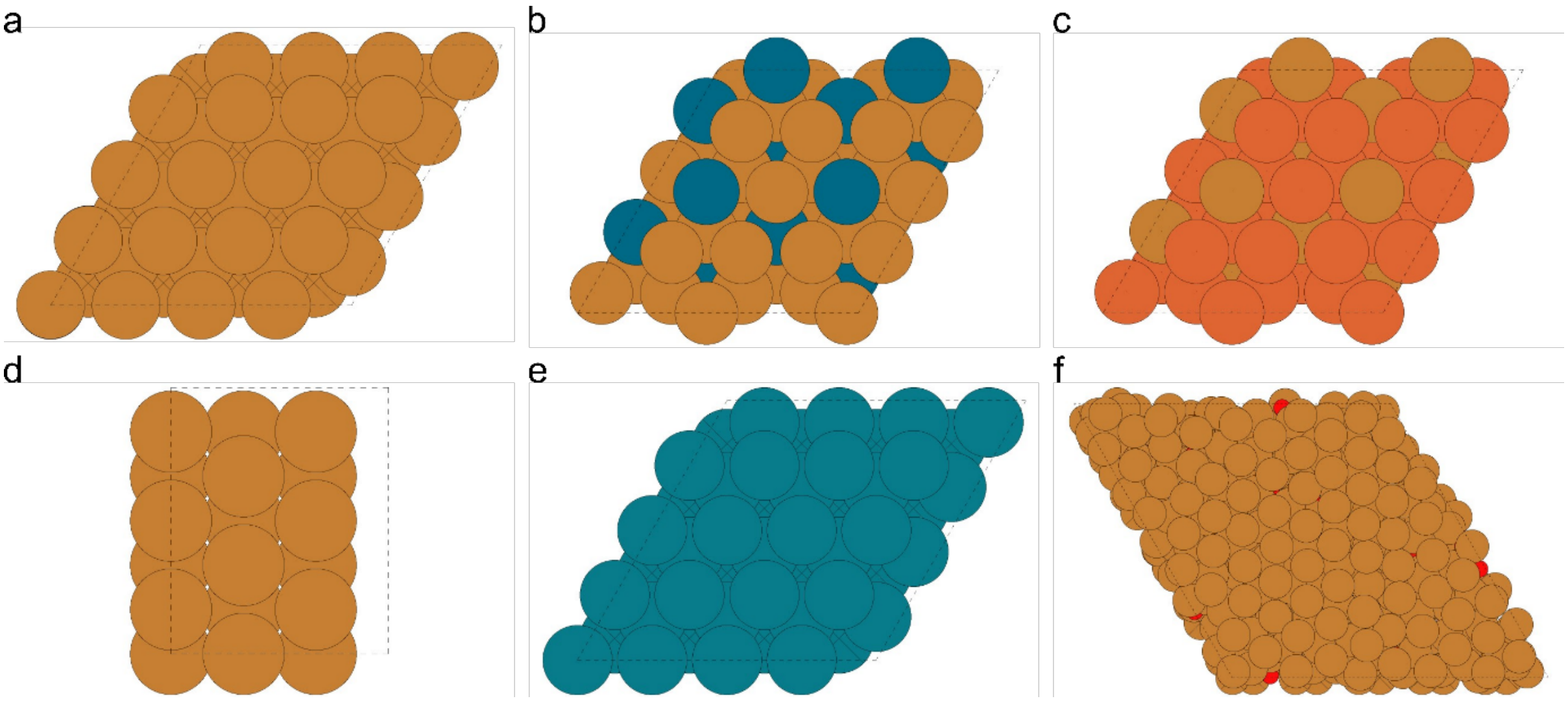

**Figure S1.** Atomic models of catalyst surfaces used for computational benchmarks and experimental case studies. (a) Cu (111), (b) $Cu_3Pd$ (111), (c) $Fe_3Cu$ (111), (d) Cu (211), (e) Rh (111) and (f) OD-Cu).

## A.2 Adsorbate Library and Placement

Twenty-four adsorbate geometries were defined by atomic offsets relative to an adsorption-site anchor (**Table S2**). Adsorbates were placed sequentially, with an inter-adsorbate minimum atomic separation of 1.5 Å under the minimum-image convention. Placements violating this cutoff were rotated around the surface normal over 12 azimuthal orientations (30° intervals), and the orientation with the largest clearance was retained. Configurations satisfying the cutoff without rotation were left unchanged. The 1.5 Å cutoff was selected to reject steric overlaps while permitting short hydrogen bonding contacts that would be excluded by larger cutoffs.

The atom indices of each adsorbate instance were retained for normal-mode attribution (**Note B.2**). Design agents also received species-specific footprint information, including atom counts, lateral extents, minimum compatible site separations, and typical neighboring site distances.

**Table S2. Adsorbate library and initial placement parameters used to construct candidate surface states. For each of the 24 adsorbate species, the table specifies the atomic anchor(s) used to position the adsorbate and the corresponding initial height(s) above the binding site.**

| Species | Anchor Atom(s) | Height (Å) | Species | Anchor Atom(s) | Height (Å) |
|---|---|---|---|---|---|
| CO | C | 1.85 | OCCHO | C, C | 1.85, 1.95 |
| COH | C | 1.85 | OHCCOH | C, C | 1.85 |
| COOH | C | 2.00 | OCCHOH | C, C | 1.85, 1.95 |
| COO | C | 2.10 | CH | C | 1.45 |
| H | H | 1.00 | $OCCH_3$ | O | 1.60 |
| OH | O | 1.80 | $CO_3$ | O, O | 1.95 |
| O | O | 1.40 | $OCHCH_2$ | O | 1.60 |
| CHO | C | 1.90 | $OCHCH_3$ | O | 1.60 |
| HCOO | O, O | 1.70 | Benzene | $C_6$ | 2.20 |
| HCOOH | O | 1.65 | Ethylidyne | C | 1.30 |
| OCCO | C, C | 1.85 | Acetylene | C, C | 1.65 |
| OCCOH | C, C | 1.85 | Propylidyne | C | 1.30 |

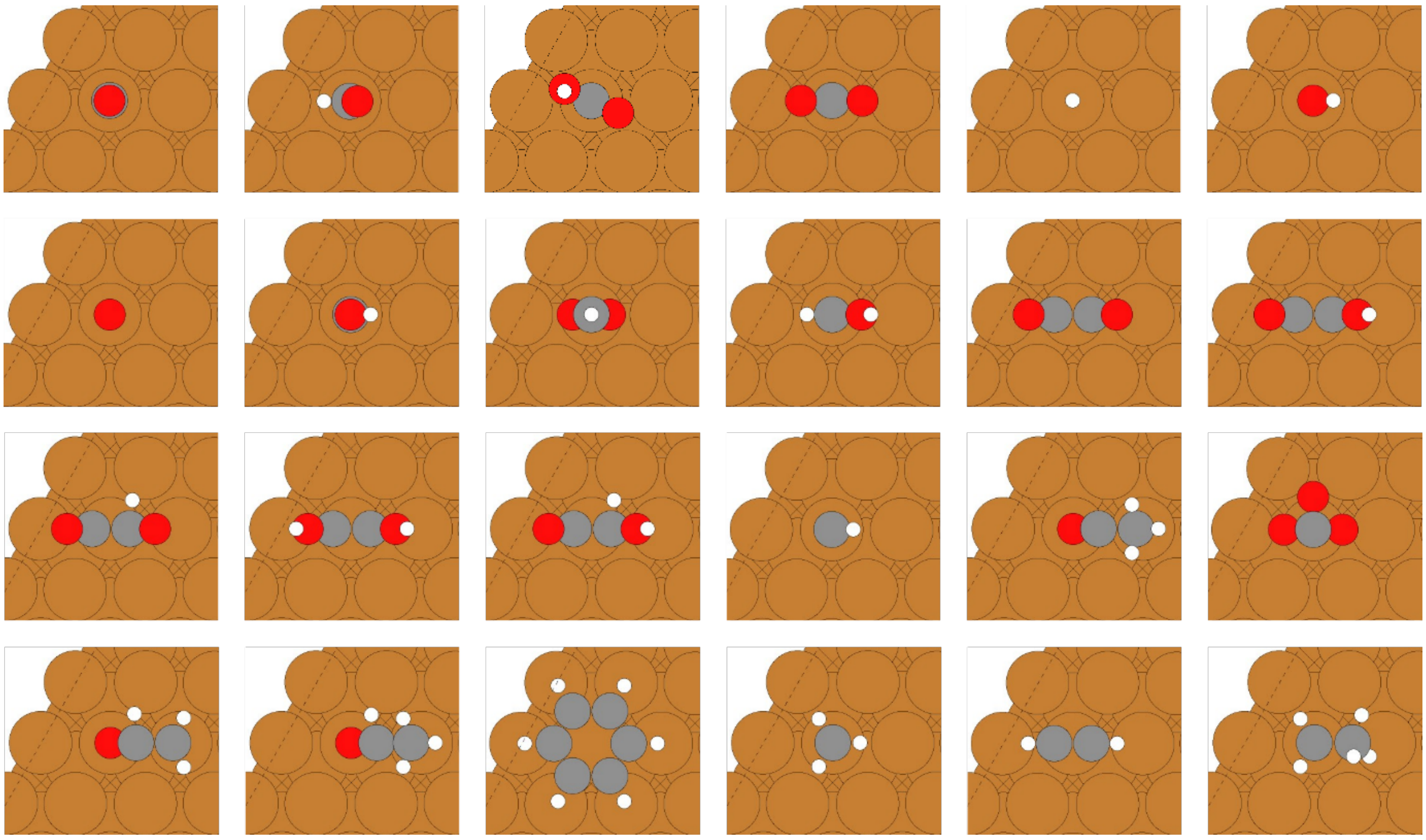

**Figure S2.** Illustrations of the 24 adsorbate species in the candidate library placed initially at atop sites of Cu (111). Structures are arranged from left to right and from top to bottom in the same order as the species listed in **Table S2**.

### A.3 Surface State Text and Arrangement Signature

Each surface state was represented by its adsorbate composition, site indices and fractional coverage, defined as the number of adsorbates divided by the number of surface atoms. An adjacency block listed connected adsorbate groups, adjacent pairs, and the neighbor count of each adsorbate (**Table S4**). Arrangement signatures were constructed from the connected components of the adsorbate induced site graph using component size, internal edge count, and the species resolved degree multiset. The signature is invariant to equivalent site indices but distinguishes different local adjacency patterns at fixed composition.

### A.4 Computational and Experimental Spectrum Representation

Calculated spectra were represented as frequency and intensity arrays. Peaks were identified after third-order Savitzky–Golay smoothing at 3, 8, and 20 $cm^{-1}$,[2,3] using a minimum prominence of 1% of the maximum intensity, minimum separation of the larger of 5 $cm^{-1}$ and the smoothing width, and minimum width of 1 $cm^{-1}$. Peaks detected within 5 $cm^{-1}$ across smoothing scales were merged by retaining the most prominent peak. Peaks from 250−4000 $cm^{-1}$ were presented to the agents along with descriptors and spectral distances calculated separately for seven frequency windows (**Table S3** and **Note D.1**).

**Table S3. Seven vibrational frequency windows used to calculate region resolved spectral distances and describe local spectral discrepancies.**

| Window | Range (cm-1) | Label |
|---|---|---|
| R1 | 250-800 | Surface-adsorbate modes, frustrated/libration. |
| R2 | 800-1300 | C–O, C–OH, O–C–O bending/skeletal modes |
| R3 | 1300-1800 | OCO stretches, carbonate/bicarbonate/formate, COOH/CHO carbonyl, $H_2O$ bend, $C_2$ intermediates |
| R4 | 1800-2250 | adsorbed CO stretch: bridge/atop CO |
| R5 | 2250-2600 | $CO_2$ asymmetric stretch / $CO_2$-derived background |
| R6 | 2600-3200 | C–H stretch + strongly H-bonded O–H |
| R7 | 3200-3800 | adsorbed OH / $H_2O$ O–H stretch, H-bonded through to free OH |

For measured targets, experimental spectra were supplied as images. Design agents received calculated peak positions and their mode attributions, whereas reflection agents additionally received seven individually normalized, region resolved spectral panels without intensity axis labels. Representative surface and spectral inputs are shown in **Table S4**.

**Table S4. Representative text-based inputs supplied to the agents for surface state inference.**

| **Example of *surface atoms*** |
|---|

Index | Symbol | Connectivity
12 | Cu | [54, 56, 58, 59, 60, 63]
13 | Cu | [55, 57, 58, 59, 61, 62]
14 | Cu | [52, 53, 54, 57, 60, 62]
15 | Cu | [52, 53, 55, 56, 61, 63]
52 | Fe | [14, 15, 54, 56, 61, 62]
53 | Fe | [14, 15, 55, 57, 60, 63]
54 | Fe | [12, 14, 52, 56, 57, 59]
55 | Fe | [13, 15, 53, 56, 57, 58]
56 | Fe | [12, 15, 52, 54, 55, 58]
57 | Fe | [13, 14, 53, 54, 55, 59]
58 | Fe | [12, 13, 55, 56, 60, 62]
59 | Fe | [12, 13, 54, 57, 61, 63]
60 | Fe | [12, 14, 53, 58, 62, 63]
62 | Fe | [13, 14, 52, 58, 60, 61]
61 | Fe | [13, 15, 52, 59, 62, 63]
63 | Fe | [12, 15, 53, 59, 60, 61]

## Example of *selectable adsorption sites*

Index | Site | Anchor atoms | Connectivity | Excludes (too close to use together)
48 | top | [48] | [49, 51, …, 117] | [64, 65, 66, 67, 68, 69]
49 | top | [49] | [48, 50, …., 121] | [64, 70, 71, 72, 73, 74]
...
63 | top | [63] | [50, 51, …, 143] | [79, 83, 99, 107, 109, 111]
64 | brg | [48, 49] | [52, 61, …, 121] | [48, 49, 66, 69, 71, 73, 112, 113]
65 | brg | [48, 51] | [55, 60, …, 127] | [48, 51, 67, 68, 81, 82, 114, 115]
...
111 | brg | [62, 63] | [50, 59, …, 142] | [62, 63, 78, 79, 106, 107, 125, 143]
112 | fcc | [48, 49, 52] | [48, 49, …, 129] | [64, 66, 71]
113 | hcp | [48, 49, 61] | [48, 49, …, 142] | [64, 69, 73]
...
143 | hcp | [59, 62, 63] | [59, 62, …, 142] | [106, 107, 111]

## Example of *adjacency block*

Adjacency: 2 adjacent pair(s) among 4 adsorbate(s), forming 2 connected group(s)
Connected groups: [CO@48, OH@49, COOH@52]; [OH@58]
Adjacent pairs: CO@48-OH@49, OH@49-COOH@52
Adjacent-neighbour count per adsorbate: CO@48: 1, OH@49: 2, COOH@52: 1, OH@58: 0
Any pair not listed as adjacent is NOT adjacent.

## Example of *arrangement signature*

1. {CHO, H, HCOO, OH} arranged as 4 connected group(s) of size 1, 1, 1, 1; 0 adjacent pair(s)
2. {CO, COOH, OH} arranged as 1 connected group(s) of size 3; 2 adjacent pair(s)
3. {CO, HCOO, OH} arranged as 3 connected group(s) of size 1, 1, 1; 0 adjacent pair(s)

## Example of *spectrum block for computational case*

Forward-simulated peaks:
1. Freq: 251.09, Intensity: 908.94
2. Freq: 274.60, Intensity: 787.25
3. Freq: 346.39, Intensity: 951.75
[…]
17. Freq: 2686.81, Intensity: 796.20
18. Freq: 3429.40, Intensity: 786.99
19. Freq: 3638.57, Intensity: 794.32
Spectrum distance to target (0 = perfect match, 1 = worst): 0.1246
Per-region distance to target (0 = identical, 1 = maximally different). Each score is 0.5 x band-position term + 0.5 x integrated-intensity term:
R1 (250–800, surface–adsorbate modes, frustrated/libration, M–CO, M–O, M–OH): 0.143 [position 0.069 = 38 cm^-1 | intensity 0.216 = 0.64x the target's integrated intensity]
R2 (800–1300, C–O, C–OH, O–C–O bending/skeletal modes): 0.135 [position 0.128 = 64 cm^-1 | intensity 0.143 = 0.75x

the target's integrated intensity]
R3 (1300–1800, OCO stretches, carbonate/bicarbonate/formate, COOH/CHO carbonyl, H2O bend, C2 intermediates): 0.152 [position 0.104 = 52 cm^-1 | intensity 0.200 = 0.67x the target's integrated intensity]
R4 (1800–2250, adsorbed CO stretch: bridge/atop CO): 0.299 [position 0.098 = 44 cm^-1 | intensity 0.500 = 0.33x the target's integrated intensity]
R5 (2250–2600, CO2 asymmetric stretch / CO2-derived background): 0.000 [silent in both — nothing to match here]
R6 (2600–3200, C–H stretch + strongly H-bonded O–H): 0.008 [position 0.015 = 9 cm^-1 | intensity 0.000 = 1.00x the target's integrated intensity]
R7 (3200–3800, adsorbed OH / H2O O–H stretch; H-bonded through to free OH): 0.011 [position 0.022 = 13 cm^-1 | intensity 0.000 = 1.00x the target's integrated intensity]

**Example of *spectrum block for experimental case***

Forward-simulated peaks:
1. 276.3 cm^-1
   calculated mode participation:
     HCOO@296 : 0.37
     CHO@289 : 0.32
     OCCO@283 : 0.30
   contributing raw modes: 252.7, 270.0, 278.0, 279.6, 299.3 cm^-1
2. 325.1 cm^-1
   calculated mode participation:
     OCCO@283 : 0.78
     CHO@289 : 0.18
     HCOO@296 : 0.03
   contributing raw modes: 299.3, 326.0 cm^-1
3. 473.9 cm^-1
   calculated mode participation:
     OCCO@283 : 0.74
     CHO@289 : 0.26
     HCOO@296 : 0.00

[…]

17. 1726.7 cm^-1
   calculated mode participation:
     OCCO@283 : 1.00
     CHO@289 : 0.00
     HCOO@296 : 0.00
18. 2890.7 cm^-1
   calculated mode participation:
     CHO@289 : 0.90
     HCOO@296 : 0.10
     OCCO@283 : 0.00
   contributing raw modes: 2888.3, 2911.6 cm^-1
19. 2908.2 cm^-1
   calculated mode participation:
     HCOO@296 : 0.87
     CHO@289 : 0.13
     OCCO@283 : 0.00
   contributing raw modes: 2888.3, 2911.6 cm^-1

## Supplementary Note B. Vibrational Models and MLFF Validation

### B.1 Machine Learning Force Field

Atomic forces were evaluated using the pretrained Universal Model for Atoms (UMA, uma-s-1p2 checkpoint, oc20 task head)[4, 5] on a single NVIDIA GPU. The same calculator was used for target and candidate evaluations.

### B.2 Forward Vibrational Spectrum Calculation and Mode Attribution

Vibrational spectra were calculated within the local harmonic approximation using adsorbate degrees of freedom. With the DFT-relaxed substrate fixed, adsorbate coordinates were optimized by BFGS to a maximum force of 0.01 eV/Å (up to 500 steps)[6]. Harmonic frequencies were obtained from the adsorbate Hessian using central finite differences (ASE Vibrations, displacement ±0.01 Å)[7]. Imaginary and nonpositive modes were excluded. The remaining frequencies were Gaussian broadened ($\sigma$ = 10 $cm^{-1}$) over 0–4000 $cm^{-1}$ on a 4000-point grid and normalized to unit maximum.

Each normal mode was assigned to adsorbate instances using the sum of mass-weighted squared atomic displacements, $\sum_i m_i|e_i|^2$. For broadened peaks, mode contributions within ±3$\sigma$ were aggregated with Gaussian weights, and the contributing raw frequencies were retained. Unassigned contributions were labeled as surface related.

### B.3 DFT Reference Calculations

DFT calculations were performed using the Vienna Ab initio Simulation Package (VASP. Version 5.4.4)[8, 9] with the projector augmented wave (PAW) method[10]. The exchange correlation energy was described using the revised Perdew-Burke-Ernzerhof (RPBE) functional within the generalized gradient approximation (GGA)[11]. A plane-wave kinetic energy cutoff of 450 eV and 3 × 3 × 1 Monkhorst-Pack k-point mesh were employed for surface calculations. Electronic occupancies were treated using first order Methfessel-Paxton smearing[12] with a width of 0.25 eV. Atomic positions were optimized while maintaining the prescribed atomic constraints and fixed lattice parameters with convergence criteria of $10^{-4}$ eV for the electronic energy and 0.05 eV/Å for the atomic forces. Subsequent force calculations for vibrational analysis employed a tighter electron convergence criterion of $10^{-7}$ eV. The resulting forces were used to obtain the reference vibrational frequencies following the procedure described in **Note B.2**.

### B.4 Validation of UMA Vibrational Frequencies

UMA vibrational frequencies were compared with DFT references for 57 single-adsorbate structures (19 species on atop, bridge, and fcc sites) and 50 randomly generated multi-adsorbate configurations on a 4 × 4 Cu (111) slab. Random configurations contained 1−6 adsorbates drawn from CO, OH, CHO, and COOH across available adsorption site types, subject to a 1.5 Å clearance criterion. The sampled adsorbate-count distribution was 20, 17, 4, 5, 2, and 2 structures with one through six adsorbates, respectively. Details are summarized in **Figure S3** and **Table S5**.

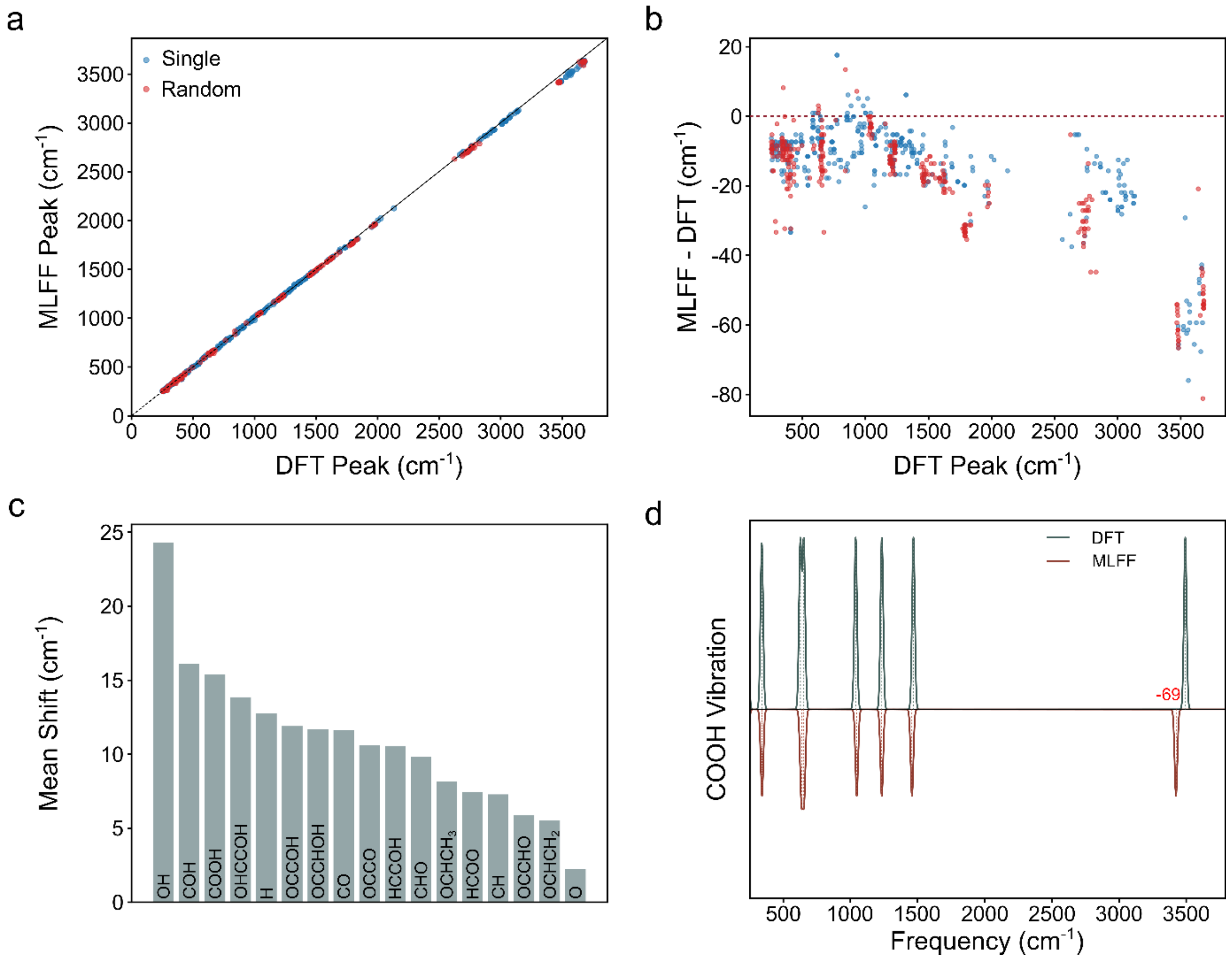

**Figure S3.** Validation of UMA-based harmonic vibrational frequencies against DFT for adsorbates on Cu (111). (a) Parity plot of matched peak positions calculated using MLFF and DFT. (b) Signed frequency deviations (MLFF − DFT) as a function of the DFT frequency. (c) Species mean absolute frequency shifts for individual adsorbates. (d) Example comparison of calculated vibrational fingerprints for COOH species.

**Table S5. Mean absolute errors (MAEs) in vibrational peak positions calculated using UMA relative to DFT on Cu (111).**

| Metrics | Single Adsorbates | Random Surface State | All |
|---|---|---|---|
| The Number of Cases | 57 | 50 | 107 |
| Peak Shift (MAE) | 10.2 | 14.9 | 12.0 |
| MAE below 3200 $cm^{-1}$ | 7.2 | 9.7 | 8.1 |
| MAE above 3200 $cm^{-1}$ | 57.6 | 56.6 | 57.1 |

## Supplementary Note C. Details of Agentic Framework

### C.1 Inference Workflow and Iterative Search

The surface state inference loop was implemented in pydantic-graph with initialization, routing, design, calculation, and reflection stages. At each round, the router selected exploration, exploitation, or Bayesian optimization (BO). Exploration and exploitation generated three candidates per round, while BO generated one. Each candidate underwent a forward spectral calculation, followed by reflection and the next routing decision.

Initialization constructed the binding site catalogue and geometric constraints, processed the target spectrum, and evaluated the initial warm-start structures. These evaluations were retained in the history without consuming the round budget. The inference history stored evaluated configurations, generation strategies, spectra, peak assignments, region-specific and overall distances where available, structural reasoning and geometric corrections. The same records supplied context to the agents, training observations to the BO surrogate, and progress statistics to the router. For calculated targets, the search ended at $D \leq 0.01$ or when the maximum number of iterations was reached (**Note D.2**). For measured targets, the router could additionally terminate the search based on its spectral assessment.

### C.2 Agent Architecture and Structured Reasoning

Five agents were implemented using pydantic-ai, exploration design, exploitation design, exploration reflection, exploitation reflection, and routing. Each used a predefined system prompt, structured outputs, and validation. A single language model was used within each run (**Table S6**).

Design agents specified adsorbate identities and binding sites, with separate reasoning fields for vibrational frequencies and spectral features. Optional target peak attributions indicated tentative spectral support for proposed adsorbates. Exploration sampled different adsorbate compositions and adjacency arrangements. Exploitation generated three independent modifications of the current best configuration, (i) one adsorbate edit, (ii) up to two coupled adsorbate edits, and (iii) rearrangement of at least two adsorbates without changing composition.

Reflection agents evaluated each candidate against the target across the seven spectral regions (**Table S3**), returned region-specific diagnoses, and identified promising candidates. The router selected the next search action from the history and reflection outputs, recording its rationale and expected outcome for comparison with the next round. For measured targets, reflection additionally returned a qualitative match score (0 to 10), a head-to-head judgment against the current best, and supporting spectral evidence (**Note F.1**).

### C.3 System Prompts

System prompts consisted of fixed agent-specific instructions and configurable blocks assembled at the start of each run. The blocks defined general spectral and binding rules, enabled species and site types, and target-specific evaluation rules. The part of prompts and instruction blocks are summarized in **Tables S6**, and the complete prompts are in https://github.com/ahrehd0506/SIREN

**Table S6. Representative system prompt excerpts for the five agents used in the SIREN workflow**

| System Prompt of Exploration Design Agent |
|---|
| You are a chemical researcher specializing in spectroscopy and catalytic<br>surface structures, working in EXPLORATION mode. Given a target spectrum,<br>propose N candidate coverage configurations whose compositions cover as<br>wide a variety of intermediate combinations as possible.<br>[...]<br>EXPLORATION AXES: COMPOSITION x ARRANGEMENT<br>A proposal is characterised by two things, and either one can be varied:<br>- Its COMPOSITION, the sorted multiset of adsorbate types; counts matter,<br>  so {CO}, {CO, CO} and {CO, CO, CO} are three compositions.<br>- Its ARRANGEMENT, which adsorbates are adjacent to which. Three mutually<br>  adjacent adsorbates, two adjacent plus a distant third, and three with no<br>  adjacencies are three DIFFERENT experiments at identical composition.<br>Two proposals are the same experiment only when composition AND adjacency<br>structure both match; moving adsorbates to other indices that reproduce the<br>same adjacency is NOT a new proposal. Every proposal must differ from every<br>other proposal in the response and from every already-evaluated entry in<br>composition, in arrangement, or in both. A composition already tried is<br>still available under a different arrangement, and that is often the |

productive move.

YOUR ROLE IS BREADTH. A separate exploitation agent refines the current best with local edits and repackings; do not duplicate that. Propose materially different hypotheses in species, counts and broad packing so the search is not trapped around one early guess. Use chemistry as a PRIOR, not as a substitute for the forward simulator.
[...]

## System Prompt of Exploitation Design Agent

You are a chemical researcher specializing in spectroscopy and catalytic surface structures, working in EXPLOITATION mode. You are refining ONE current best-so-far coverage for a single target spectrum.
[...]
THREE-ARM EXPLOITATION BATCH
All three proposals are independent alternatives starting from the SAME reference; never chain proposal 1 from proposal 0 or 2 from 1.

PROPOSAL 0 — STRICT LOCAL: ONE ADSORBATE-LEVEL EDIT
Exactly ONE adsorbate-level intervention relative to the reference: MOVE one adsorbate to another site, ADD one, REMOVE one, or REPLACE one by another species. Everything else stays exactly unchanged. It tests one sharply defined explanation of the residual mis-fit.

PROPOSAL 1 — LOCAL/COUPLED: AT MOST TWO ADSORBATE-LEVEL EDITS
At most TWO adsorbate-level interventions. Prefer a second single edit; use two coupled edits only when the hypothesis itself requires coordination (creating or breaking a specific adjacency, separating a coupled pair, jointly relocating two adsorbates), and say why one edit would not test it.

PROPOSAL 2 — FIXED-COMPOSITION REPACKING: SAME SPECIES, >=2 ADSORBATES MOVED
Keep the species multiset and total count EXACTLY as in the reference, but move at least two adsorbates to test a qualitatively different packing / adjacency hypothesis (clustered -> dispersed, adjacent pair -> separated, isolated OH -> H-bond-capable neighbourhood, one group -> several groups). Do NOT add, remove or replace species here; a large rearrangement is deliberately allowed. Proposal 2 must not collapse into another single-edit candidate.

HOW TO CHOOSE
- Start from the OBSERVED per-region errors of the reference.
- Use previously evaluated configurations as controlled experiments:
  observed intervention -> spectral change outweighs a verbal assignment.
- Preserve a successful structural core in proposals 0 and 1 unless the
  data directly implicate it; proposals 0 and 1 should test different
  explanations rather than two trivial variants of one site move.
- Proposal 2 must test a NAMED packing hypothesis, not random rotation.
[...]

## System Prompt of Exploration Reflection Agent

You are a chemical researcher reviewing EXPLORATION proposals. The design agent produced N candidate coverages for a target spectrum. Describe how each proposal's simulated spectrum differs from the target region by region, and shortlist the most promising.

You are descriptive, not prescriptive: do NOT recommend what the design agent should try next, and you do not decide what the search does next — a routing agent reads your analysis and chooses. Report what you observe.

[...]
OUTPUT
1. diagnoses: for each proposal, one analysis per region (R1, R2, R3, R4, R5, R6, R7), 1–2 sentences each, citing concrete band positions and which band groups are present, absent or displaced.
2. best_candidate_indices: a shortlist of 1–3 proposals. Your job is to filter out the obviously wrong ones.
3. overall_diagnosis: up to 3 sentences on which composition directions look promising and which are clearly off.
[...]
Ground every analysis in actual differences; no generic feedback. Keep structural causes uncertain: count, site/environment, adjacency/coupling and modes shifting across region boundaries can all change a region.
[...]

## System Prompt of Exploitation Reflection Agent

You are a chemical researcher reviewing ONE evaluated coverage. Describe how its simulated spectrum differs from the target region by region, and decide whether the two match within tolerance.

The candidate's "Proposal origin" is exploitation (refined from the best so far by a reasoning agent, with its reasoning attached) or bo (picked by a Gaussian-process surrogate; no reasoning is attached and none is expected — judge it on the spectra alone).

You are descriptive, not prescriptive: do NOT recommend what the design agent should try next, and you do not decide what the search does next — a routing agent reads your analysis and chooses. Report what you observe.
[...]
OUTPUT
1. spectra_match: whether the candidate's spectrum is consistent with the target within tolerance.
2. region_diagnoses: one analysis per region (R1, R2, R3, R4, R5, R6, R7), 1–2 sentences each, citing concrete band positions and which band groups are present, absent or displaced.
3. overall_diagnosis: up to 3 sentences on the overall match.
[...]
Ground every analysis in actual differences; no generic feedback.
[...]

## System Prompt of Router Agent

You are the search controller for a spectrum-inversion loop. Each round one action is taken, its forward simulations are run, and a reflection agent describes how the result compares to the target. Pick the NEXT action purely on which one addresses what is currently wrong with the fit.

ACTIONS
- exploration: BREADTH. A reasoning agent proposes coverages with adsorbate types, counts and packings nothing has tested yet. Strong when few combinations have been tried, or a whole region of the best coverage is badly mis-fit in POSITION, which means the types themselves are wrong.
- exploitation: DEPTH. From the current best, a three-candidate batch: (0) one strict local edit, (1) a local or coupled <=2-adsorbate edit, (2) a fixed-composition repacking. Strong when a competitive composition exists and the question is whether a local correction or the collective packing is limiting the fit.
- bo: a Gaussian-process surrogate proposes a point. Judge it by its EMPIRICAL recent performance; if recent BO proposals are repeatedly much

```
  worse than the best, deprioritise it until the data have changed.
- stop: end the run. Only when the spectra genuinely match — never
  because progress slowed.

HARD CONSTRAINT: choose only from the ALLOWED actions listed in the prompt.

HOW TO WEIGH THE SIGNALS
- Rounds since the best last improved is a staleness signal, but also
  inspect WHICH interventions were tried.
- Per-region distances say WHAT is wrong, not uniquely WHY. A large
  POSITION mismatch can come from type, site, environment or coupling;
  prefer exploration when it persists across several local interventions
  or several regions jointly contradict the composition. A large INTENSITY
  mismatch can come from count, site, coupling, hydrogen bonding or a shift
  across a region boundary; prefer a count-changing exploitation step only
  when evaluated local comparisons support it.
- Exploitation proposals 0 and 1 are causal clues (controlled intervention
  -> change); proposal 2 tests whether collective packing matters. If local
  probes keep improving, stay local; if repacking improves at fixed
  composition, keep the composition and exploit arrangement; if all three
  arms fail, favour exploration of new compositions over rebuilding the
  same best.
- Repeatedly choosing an action that underdelivers against its stated
  expectation is the main failure mode to avoid.
[...]
```

## C.4 Agent Input and Contextual Information

Agent inputs were rebuilt after each round, combining a fixed system description with the target spectrum, selected evaluations, recent reflections and the current task. Design prompts included catalyst metadata, binding site information, and adsorbate footprints. Reflection prompts included the relevant evaluated configurations.

For calculated targets, prompts contained numbered vibrational peaks and numerical descriptors for R1–R7. For measured targets, the spectrum was supplied as images without assigned peaks or numerical distances. The agents compared band presence, ordering, and spacing rather than absolute intensities or exact frequencies. Experimental measurement notes can be included to describe details of spectrum images (**Table S13**).

## C.5 Inference Memory and Feedback Integration

All evaluated configurations and reflection outputs were retained in the inference history. Design prompts included warm-start observations and up to 15 selected prior configurations, ranked by the sum of regional distances for calculated targets or by qualitative match score for measured targets.

Recent feedback was restricted to the three latest relevant rounds, exploration feedback for exploration agents, exploitation and BO feedback for exploitation agents. The exploration agent additionally received previously tested composition and arrangement combinations. The exploitation agent received previously evaluated placements within two adsorbate edits of the current best or with identical composition. The router received summaries of search progress and outcomes. The complete history, rather than the truncated prompt context, was used to train the BO surrogate.

## Supplementary Note D. Spectral Distance Metric and Convergence Criterion

### D.1 Definition of Spectral Distance

The objective combined band-distribution and integrated-intensity differences across seven frequency windows (**Table S3**). Spectral intensities were clipped at zero before regional integration.

For two spectra, *a* and *b*, the non-negative spectral intensities were defined as $I_a^+(\nu) = \max[I_a(\nu),0]$ and $I_b^+(\nu) = \max[I_b(\nu),0]$. The integrated intensities within a frequency region $w = [\nu_1,\nu_2]$ were calculated as

$$A_w = \int_{\nu_1}^{\nu_2} I_a^+(\nu)\, d\nu$$
$$B_w = \int_{\nu_1}^{\nu_2} I_b^+(\nu)\, d\nu \qquad \text{(D1)}$$

To quantify differences in spectral band position and distributions, we employed the one-dimensional Wasserstein distance ($W_1$), also known as the earth mover's distance. The intensity distributions within each region were first normalized to unit integrated intensity, yielding $\hat{I}_{a,w}$ and $\hat{I}_{b,w}$. The frequency-dependent distance was then defined as

$$d_w^{shape} = \min\left(\frac{W_1\left(\hat{I}_{a,w},\hat{I}_{b,w}\right)}{\nu_2 - \nu_1}, 1\right) \qquad \text{(D2)}$$

The shape term was the one-dimensional Wasserstein distance between unit area spectra, divided by the window width. The unnormalized distance ($cm^{-1}$) was also included in agent feedback. The intensity term compared regional integrated areas to retain information discarded by unit-area normalization,

$$d_w^{intensity} = \frac{\left|A_w - B_w\right|}{A_w + B_w} \qquad \text{(D3)}$$

The regional distance was the equally weighted sum of the shape and intensity terms,

$$d_w = \frac{1}{2} d_w^{shape} + \frac{1}{2} d_w^{intensity} \qquad \text{(D4)}$$

To avoid numerical artifact arising from Gaussian broadening in frequency regions without meaningful vibrational features, a region was considered spectrally inactive when its integrated intensity was below $10^{-6}$ times the maximum regional integrated intensity across the two spectra. Regions inactive in both spectra were excluded from the evaluation, whereas regions inactive in only one spectrum were assigned a distance of $d_w = 1$, representing the presence of a spectral feature in only one spectrum.

The overall spectral distance was obtained by averaging the regional distances over the set of active regions, $W_{active}$

$$D(a, b) = \frac{1}{|W_{active}|} \sum_{w \in W_{active}} d_w \quad \text{(D5)}$$

The mean regional distance was used as the search objective and regional components were provided to the agents for feedback.

**D.2 Calibration of Convergence Threshold**

A convergence threshold was calibrated to distinguish configurations with matching composition and local adjacency from structurally non-equivalent configurations that may have similar spectra. For Cu (111), Cu (211), $Cu_3Pd$ (111) and $Fe_3Cu$ (111), 5000 random configurations per surface were generated using CO, CHO, COOH, and OH, subject to a 1.5 Å clearance cutoff. Configurations contained 1–5 adsorbates, except Cu (211), for which 1–3 were allowed.

Pairs were grouped by composition and adjacency signature, equivalent arrangements, identical composition with different adjacency or different composition. Fifteen pairs from each group per surface were selected (180 pairs, 354 distinct structures). Pairwise spectral distances were calculated using **Note D.1**. The smallest distance between non-equivalent configurations was 0.0108 (**Figure S4**, **Tables S7–S8**). A threshold of $D \leq 0.01$ was adopted, accepting 23 of 60 equivalent pairs and no non-equivalent pairs in the calibration set (100% observed precision).

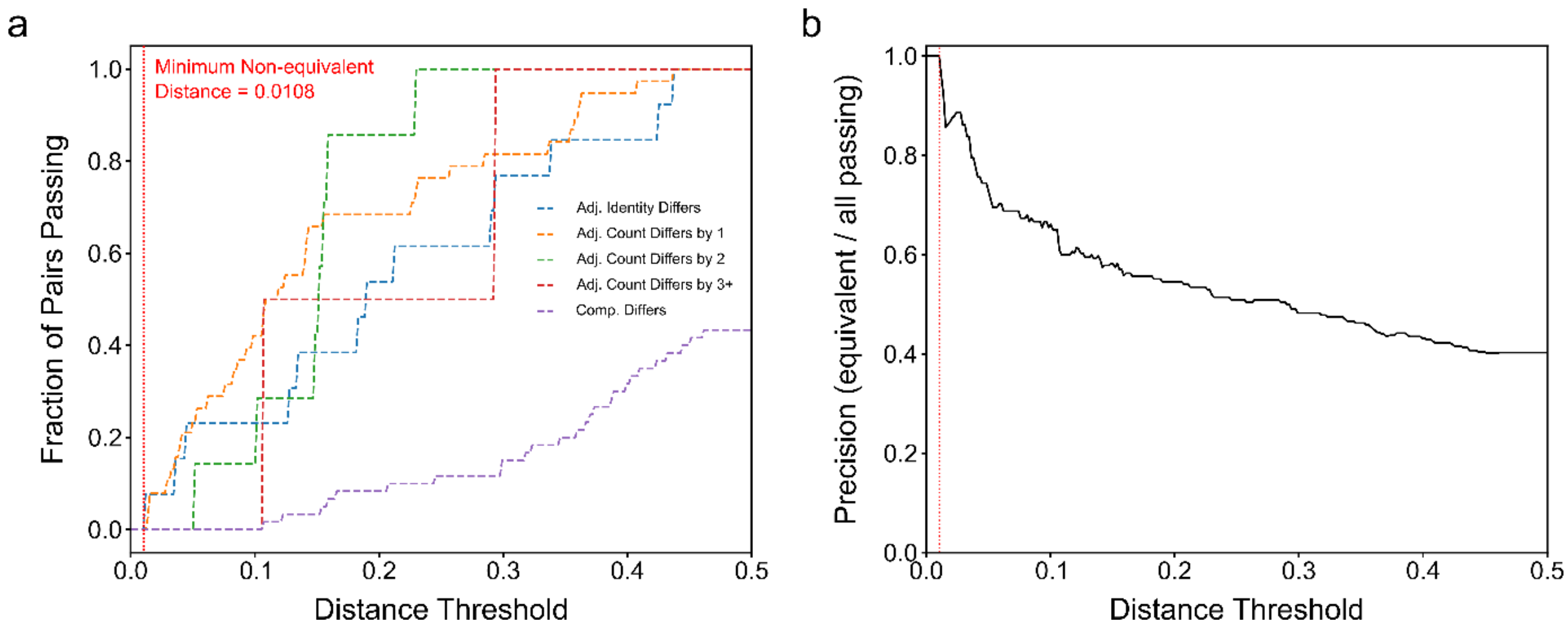

**Figure S4.** Calibration of the spectral distance threshold for recovery of a target vibrational fingerprint. (a) Fraction of configuration pairs passing a given distance threshold. (b) Precision for identifying structurally equivalent configuration pairs among all pairs passing the threshold. The vertical red dashed line indicates the adopted criterion, D = 0.01.

**Table S7.** Distribution of pairwise vibrational spectral distances used to calibrate the convergence criterion.

| Pair Class | N | Min Distance | Median Distance |
|---|---|---|---|
| Equivalent | 60 | 0.0000 | 0.0229 |
| Adjacency Differ | 13 | 0.0108 | 0.1893 |
| Adjacent Pair Differ by 1 | 38 | 0.0130 | 0.1135 |
| Adjacent Pair Differ by 2 | 7 | 0.0515 | 0.1505 |
| Adjacent Pair Differ by 3 | 2 | 0.1061 | 0.2000 |
| Different Composition | 60 | 0.1069 | 0.5362 |

**Table S8.** Classification of structurally equivalent surface state pairs at different spectral distance thresholds

| Threshold | Equivalent Pairs Accepted | Non-equivalent Pairs Accepted | Precision |
|---|---|---|---|
| 0.005 | 22 | 0 | 100% |
| 0.010 | 23 | 0 | 100% |
| 0.015 | 24 | 4 | 86% |
| 0.020 | 27 | 4 | 87% |

**Supplementary Note E. Computational Benchmark**

**Table S9.** Experimental design and search settings for the computational benchmarks.

| Setting | Value |
|---|---|
| Surface | Cu (111), Cu (211), $Cu_3Pd$ (111), $Fe_3Cu$ (111) |
| LLM Model | GPT-5-mini, GPT-5.6-luna, GPT-6-luna |
| Router Policy | Exploration, Exploitation, BO |
| Runs | 10 on each surface |
| Rounds | 50 on each run |
| Warm start | CO, OH, COOH, CHO each alone on one top site |
| Few-shot Window | 15 + warm starts |

**Table S10.** Average computational and language model usage for complete 50 round SIREN benchmark runs. Averages were obtained from 28 gpt-5-mini runs and 21 gpt-5.6-luna runs that reached the $50^{th}$ round without early termination.

| Model | Component | Calls | Wall Time (s) | Input Tokens | Output Tokens |
|---|---|---|---|---|---|
| gpt-5-mini | Spectrum Calculation | 149 | 3,240 | - | - |
| | Exploit. Design Agent | 27.4 | 2,166 | 797,793 | 214,230 |
| | Explor. Design Agent | 22.2 | 1,821 | 789,255 | 170,093 |
| | Exploit. Reflection Agent | 82.6 | 2,179 | 1,895,063 | 205,150 |
| | Explor. Reflection Agent | 22.2 | 832 | 496,459 | 79,866 |
| | Router Agent | 50.0 | 727 | 206,653 | 68,039 |
| | Total | 353 | 10,966 | 4,185,233 | 737,379 |
| gpt-5.6-luna | Spectrum Calculation | 148 | 3,229 | - | - |
| | Exploit. Design Agent | 27.4 | 1,058 | 1,625,554 | 106,216 |
| | Explor. Design Agent | 21.9 | 908 | 1,133,554 | 81,631 |
| | Exploit. Reflection Agent | 82.9 | 698 | 1,782,732 | 78,624 |
| | Explor. Reflection Agent | 21.9 | 359 | 501,710 | 41,902 |
| | Router Agent | 50.0 | 316 | 201,046 | 17,172 |
| | Total | 352 | 6,569 | 5,244,595 | 325,545 |

### E.1 Generation of Computational Target Structures

Computational targets were generated by first sampling an adsorbate count, followed by sequential assignment of CO, OH, CHO, or COOH to shuffled atop sites. Adsorbates were sampled with replacement and placed subject to the 1.5 Å clearance criterion, including azimuthal adjustment; rejected placements could reduce the realized count. Target configurations were retained as hidden references, and their calculated spectra alone were supplied to the agents. Benchmark settings are listed in **Table S9**.

### E.2 Bayesian Optimization Baseline

The BO-only baseline used the same targets, warm-start evaluations, surface models, placement constraints, spectral calculator, round budget, and convergence threshold as the agent-guided runs[13, 14]. It omitted LLM-based design, reflection, and routing. A count-first sampler was used for BO initialization to avoid overrepresenting densely occupied structures under uniform categorical site sampling. The sampler selected the adsorbate count before assigning species and sites.

The BO surrogate was initialized using four warm-start observations and four random configurations, followed by sequential candidate selection and retraining[15, 16]. Within the agent framework, BO was also available as a routing action and was fitted using all accumulated observations. Settings are summarized in **Table S11**.

**Table S11.** Configuration of the BO baseline used for computational benchmarks.

| Setting | Value |
| --- | --- |
| Library | Scikit-optimize 0.10 Optimizer (base_estimator=GP, acq_func=EI, n_initial_points=0, random_state=42) |
| Search Space | 25 options × 16 sites on the (111) slabs, × 9 on (211) |
| Training Data | every observation in the history, including warm starts |
| Initial phase | 4 warm starts + 4 random configurations |
| Batch size | 1 |
| Round | 50 |

## Supplementary Note F. Experimental Evaluation and Post-hoc Analysis

### F.1 Experimental Application Protocol

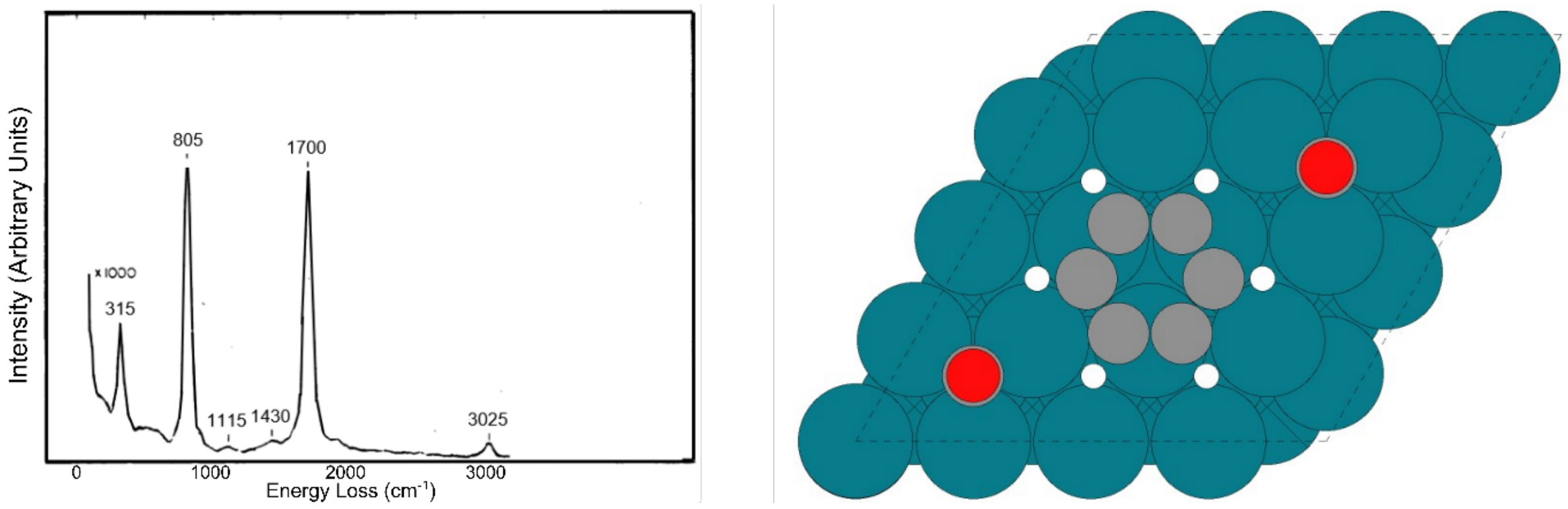

**Figure S5.** Experimental HREELS reference spectrum and the corresponding benzene and two CO co-adsorbate model on Rh (111)

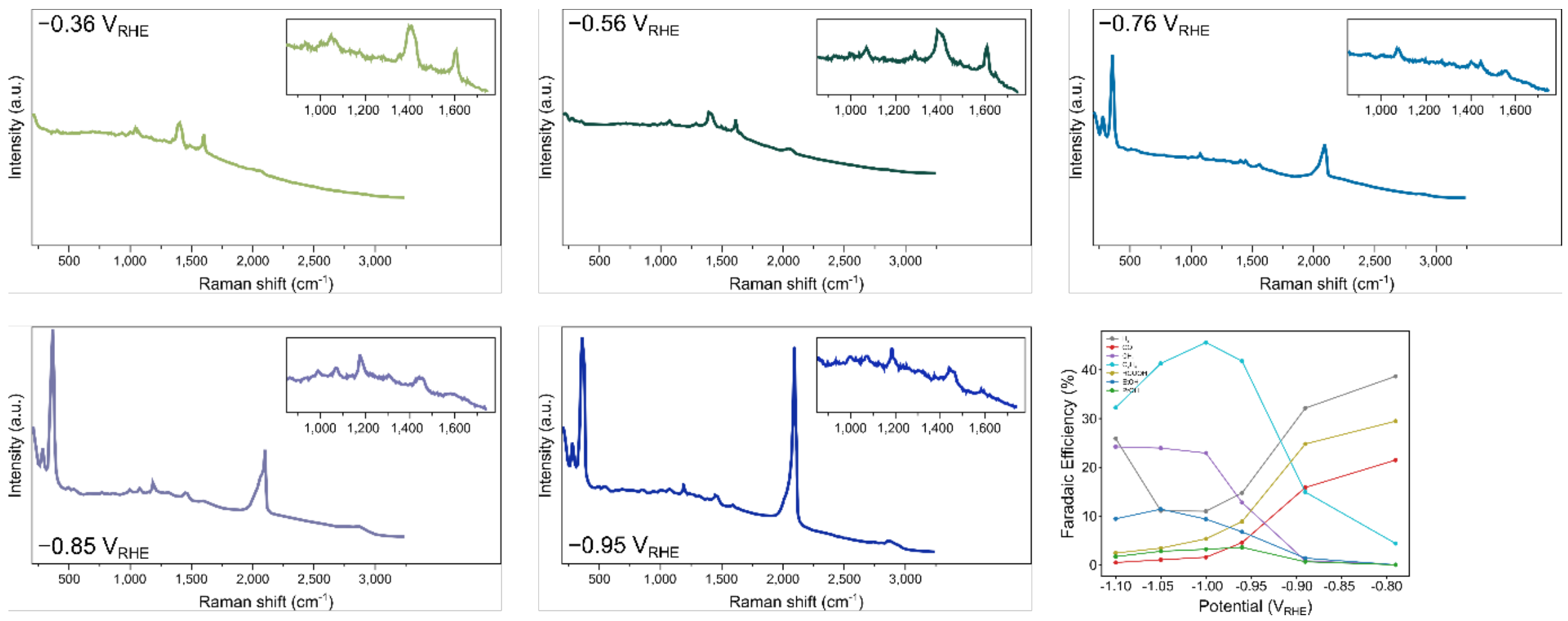

**Figure S6.** Experimental information used for the OD-Cu $CO_2$ electroreduction case study. Five panels show literature-reported in situ surface-enhanced Raman spectroscopy (SERS) spectra measured at −0.36, −0.56, −0.76, −0.85, and −0.95 $V_{RHE}$. The remaining panel shows published product Faradaic efficiencies as a function of applied potential

**Table S12.** Experimental inputs and candidate search spaces for the Rh (111) HREELS and OD-Cu SERS case studies.

| System | Measurement | Slab | Site | Adsorbates Pool |
|---|---|---|---|---|
| Rh (111) | specular HREELS, 300 K, 0–3200 $cm^{-1}$, low-frequency segment magnified ×1000 | Rh (111) 4×4 | 96 sites | CO, CHO, COH, COOH, OH, H, O, Benzene, Ethylidyne, |
| | | | | Acetylene, Propylidyne |
| OD-Cu, $CO_2RR$ | in situ SERS, electrochemically treated Cu, $CO_2$-saturated 0.1 M $NaClO_4$, −0.36, −0.56, −0.76, −0.85, −0.95 V vs RHE (main panel + 900–1700 $cm^{-1}$ inset) | $Cu_{444}O_{48}$ reconstructed OD-Cu model | 47 sites (17 top, 22 bridge, 8 fcc) | CO, OCCO, OCCOH, $OCHCH_2$, $OCHCH_3$, OCCHO, OHCCOH, OCCHOH, CHO, CH, $CO_3$, HCOO, HCOOH (13) |

Experimental cases used measured spectral images, catalyst models[17, 18], selectable adsorption sites, species pools, and measurement notes (**Figure S5** and **Tables S12–S13**). Simulated and measured spectra were compared by band group presence, order and spacing, without a numerical spectral distance objective. Each reflection assigned a qualitative match score from 0 to 10 and a direct comparison with the current best, supported by region-specific spectral evidence. Absolute intensities, linewidths and spectral densities were not used for judgment[19].

The first evaluated candidate became the current best. In later rounds, only the highest scored candidate from the round (selected from the exploration shortlist when applicable) challenged the current best. Replacement followed the reflection agent's direct better-than-current best judgment rather than the score alone.

### F.2 Post-hoc Analysis of Inferred Surface States

For each potential, final current best configurations from independent runs were pooled to obtain adsorbate occurrence fractions. A separate two-stage LLM analysis examined the inferred surface states (**Table S14**). First, potential-dependent mechanistic hypotheses were generated from inferred species and their co-occurrence without product and selectivity data. Second, literature Faradaic efficiency data were introduced to compare the hypotheses with observed product trends and identify agreements, discrepancies, and alternative explanations[20-23].

**Table S13.** Experimental notes supplied to the SIREN agents during evaluation of measured spectra.

| System | Experimental notes |
| --- | --- |
| Rh (111) | Experimental specular high-resolution electron energy loss spectroscopy (HREELS) vibrational spectrum measured on a Rh (111) surface at 300 K. The horizontal axis is vibrational energy expressed as wavenumber (cm^-1), increasing from left to right. The vertical axis is relative HREELS intensity in arbitrary units.<br>Treat peak positions, ordering, spacing, multiplicity, and the presence or absence of bands as the primary evidence. Absolute experimental intensities are not directly comparable with the calculated phonon-DOS intensity because HREELS intensities depend on electron-scattering cross sections and surface selection rules.<br><br>The spectrum covers roughly 0 to 3200 cm^-1 and nothing above that was recorded. If a candidate places a band above ~3200 cm^-1, that is NOT evidence against the candidate: the measurement simply does not reach there. Do not penalise a candidate for a band in a window the measurement does not cover, and do not treat that window as empty.<br><br>The low-frequency part of the trace carries a "x1000" magnification marker, so the band drawn near 315 cm^-1 is far weaker in reality than its plotted height suggests — of order a thousand times weaker than the bands drawn at comparable height further right. The exact segment the marker covers is not fully clear from the figure. Therefore use the low-frequency band as evidence that something absorbs there, not as a height to compare against bands in other windows. |
| OD-Cu | In situ surface-enhanced Raman spectroscopy (SERS) spectrum acquired during CO2 electroreduction on an electrochemically treated Cu electrode at {U} V vs RHE in CO2-saturated 0.1 M NaClO4 aqueous electrolyte.<br>The computational surface is an oxide-derived Cu (OD-Cu) model representing the reconstructed Cu surface under reaction conditions. The main panel shows the broad Raman spectral range, while the inset enlarges the 900-1700 cm^-1 region to resolve weaker vibrational features.<br><br>Infer a chemically plausible adsorbate configuration that jointly explains the major bands across the full spectrum. Compare primarily the presence, number, ordering, spacing, and overall pattern of vibrational bands. Absolute SERS intensities are not directly comparable with calculated harmonic vibrational fingerprints, and moderate systematic frequency shifts between experiment and calculation are expected. |

**Table S14.** Prompts used in the two stage post-hoc mechanistic analysis of inferred OD Cu surface states.

| Prompt Stage | Prompt |
| --- | --- |
| System Prompt | You are an expert in electrochemical CO2 reduction on copper-based catalysts, surface vibrational spectroscopy, and heterogeneous reaction mechanisms.<br><br>Your task is to extract chemically defensible mechanistic insights from surface states inferred by an AI spectrum-analysis agent.<br><br>Prioritize chemical validity, evidence-based reasoning, and critical evaluation over constructing a coherent or favorable narrative.<br><br>Treat inferred surface species as tentative assignments rather than experimentally confirmed intermediates. |
| | Distinguish among:<br>- Evidence directly supported by the inferred states.<br>- Mechanistic hypotheses consistent with established CO2RR chemistry.<br>- Speculative pathways requiring additional validation.<br><br>Do not fabricate literature references, reaction barriers, rate constants, surface coverages, or quantitative product selectivities. |
| Inferred States | SYSTEM: Oxide-derived Cu catalyst, electrochemical CO2 reduction toward C2+ products, investigated using operando SERS. An AI spectrum agent independently reconstructed plausible adsorbate configurations from the SERS spectrum at several applied potentials.<br><br>INFERRED SURFACE STATES:<br>{for each potential: "-U V vs RHE \| n runs \| per run: CO3+HCOO ; ... \| pooled species count: {...}"}<br><br>TASK: Analyze the inferred surface states to identify potential-dependent reaction pathways and mechanistic insights.<br><br>1. Surface-state analysis — recurring vs potential-dependent adsorbates; consistently inferred vs isolated assignments; co-occurrence within runs; run frequency = inference reproducibility, not coverage; intermediate vs spectator vs ambiguous assignment; absence from the selected configuration is not proof of physical absence.<br><br>2. Reaction pathway inference — plausible pathways to CO, formate, methane, ethylene, ethanol and other C2+ products; role of each intermediate; C1 hydrogenation, C–C coupling, oxygen retention, deoxygenation, competing pathways; shared vs discriminating intermediates; OCHCH2/OCCO/OCCHO/OCCOH/OCHCH3.<br><br>3. Potential-dependent interpretation — emergence, persistence, suppression; occurrence vs rate/selectivity; robust vs assignment-sensitive trends.<br><br>4. Mechanistic insight — connected/competing pathways, contradictions, testable hypotheses, evidence required.<br>OUTPUT: A. concise analysis per potential; B. table Potential \| Key surface evidence \| Plausible pathways \| Supporting evidence \| Counterevidence \| Confidence; C. potential-dependent reaction network; D. three to five key insights beyond restating the species; E. limitations and alternative explanations. Confidence must be qualitative and justified. Do not use experimental Faradaic efficiencies at this stage. |
| Comparison with FE | SYSTEM: Oxide-derived Cu catalyst, electrochemical CO2 reduction. The attached image presents literature-reported Faradaic efficiencies for CO2RR products at several applied potentials. An independent spectrum-based mechanistic analysis was performed without access to the FE data.<br><br>INFERRED SURFACE STATES: {as in stage 1}<br>PREVIOUS MECHANISTIC ANALYSIS: {stage-1 output}<br><br>TASK: Critically compare the inferred surface chemistry and proposed reaction pathways against the product selectivity shown in the FE figure.<br>1. Experimental FE interpretation (read labels, potentials, trends; approximate values only where visually distinguishable; do not force FE to sum to 100 %).<br><br>2. Potential alignment (distinguish potentials with both data types, surface-state only, FE only; compare neighbouring potentials qualitatively; no interpolation of adsorbate populations).<br><br>3. Intermediate–product relationships (detection of an intermediate does not imply high FE or rate; separate C2 formation from ethylene-vs-ethanol selectivity; multi-pathway |

intermediates are non-unique).

4. Consistency and discrepancies (agreements, mismatches, candidate causes: coverage, stability, kinetics, competing pathways, spectroscopic ambiguity, conditions; do not revise a hypothesis solely to agree with FE).

5. New mechanistic insights, testable hypotheses, unresolved questions.
IMPORTANT LIMITATIONS: FE data are from another publication and are contextual evidence, not validation unless conditions are shown comparable; run frequency is not coverage; inferred abundance is not flux; CO3/HCO3 species may be electrolyte/spectator chemistry; HER may give H2 FE without an adsorbate assignment; flag conflicts with established chemistry rather than forcing agreement.

OUTPUT: A. potential-by-potential analysis; B. table Potential | Surface-state evidence | Major FE products | Mechanistic interpretation | Agreement/discrepancy | Limitations; C. pathway interpretation distinguishing supported / plausible / speculative; D. three insights, each with Observation, Mechanistic interpretation, Alternative explanation, Evidence needed; E. a 150–200 word discussion paragraph.

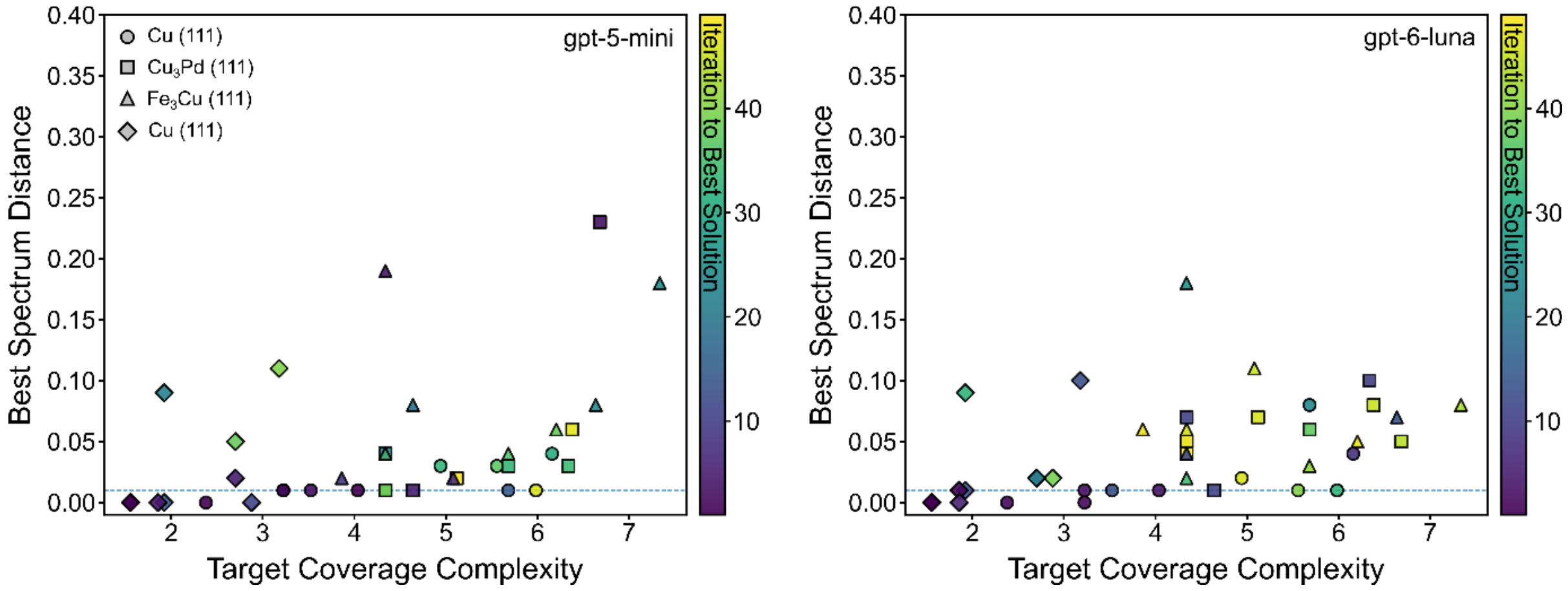

**Figure S7.** Additional language model results for the computational benchmark.

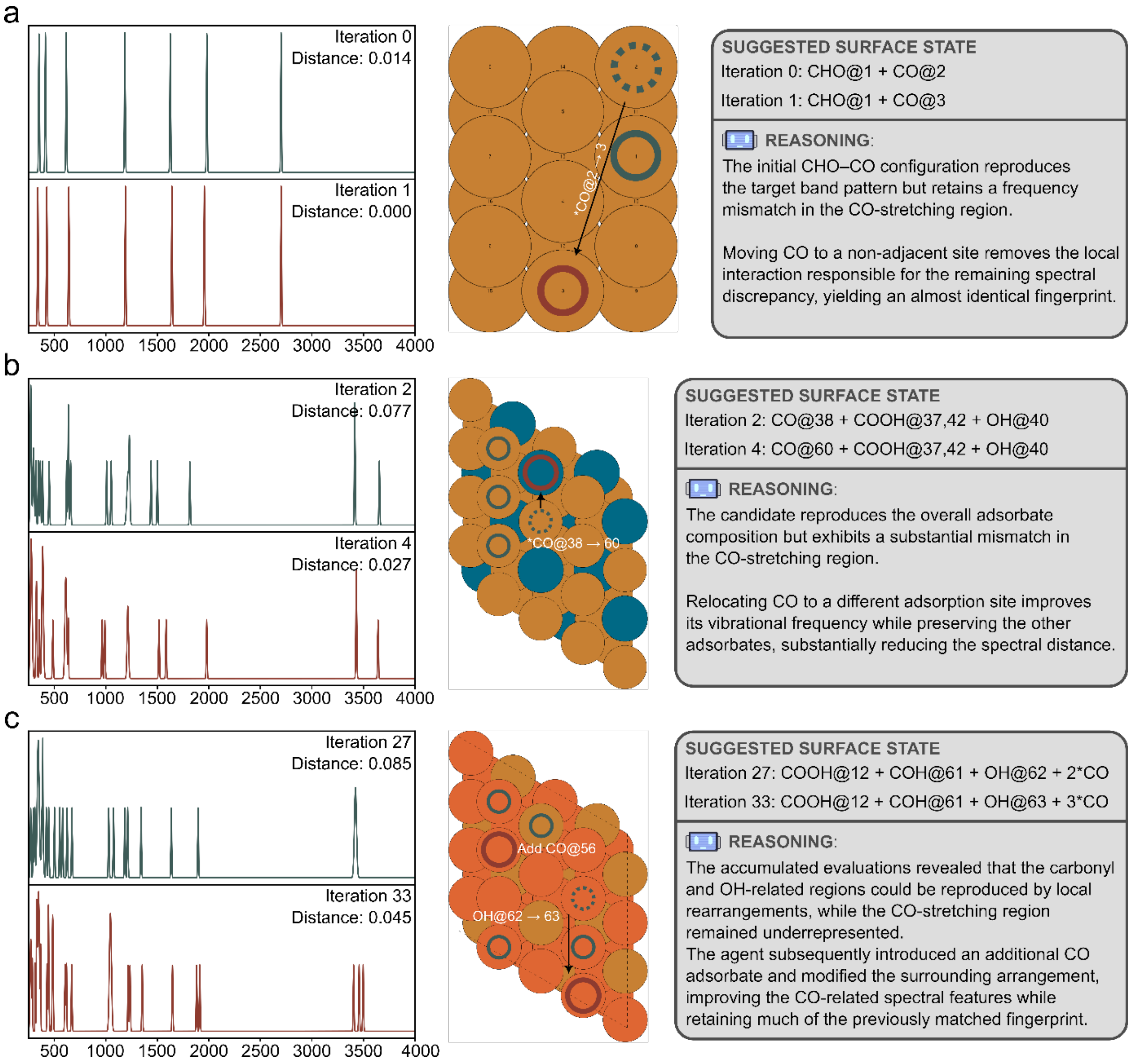

**Figure S8**. Additional examples of chemically guided structural modifications during SIREN inference. (a) Relocation of CO away from a neighboring CHO adsorbate without changing adsorbate composition. (b) Reassignment of the CO binding site, while retaining the other adsorbates, improves the CO stretching region. (c) Addition of CO and relocation of OH using previous feedback.

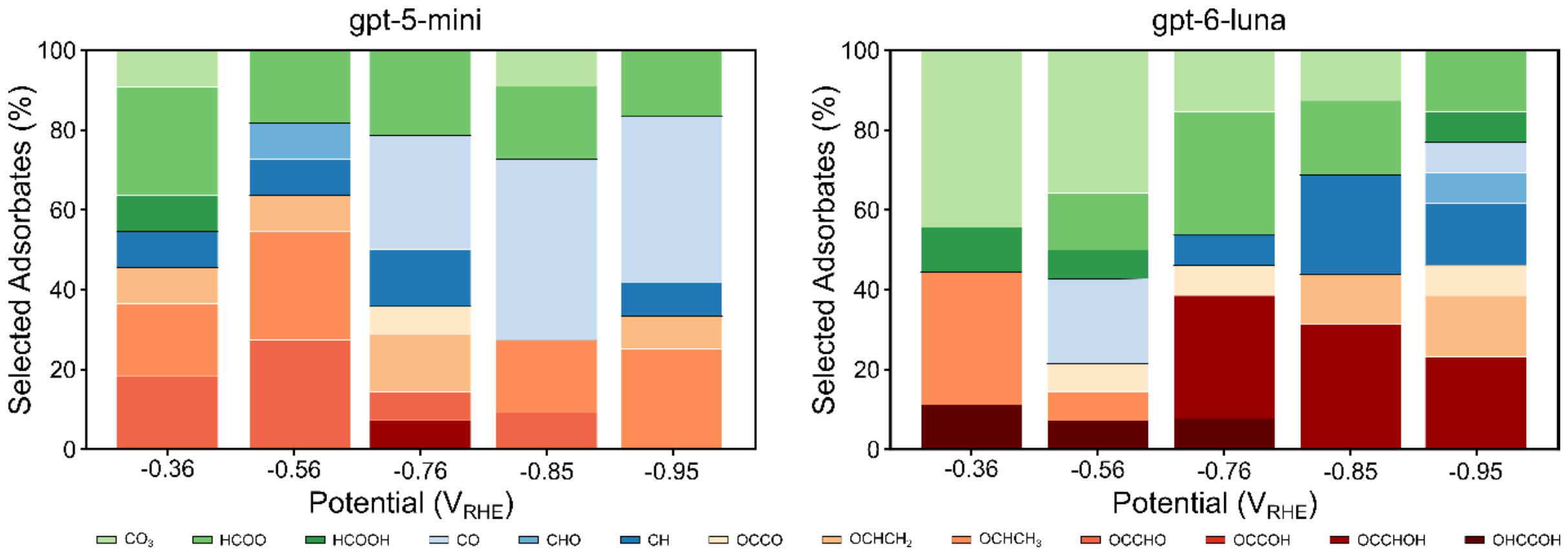

**Figure S9**. Fractions of inferred adsorbates at each potential, obtained by pooling adsorbate counts from the final selected configurations of five independent runs, from gpt-5-mini and gpt-6-luna.

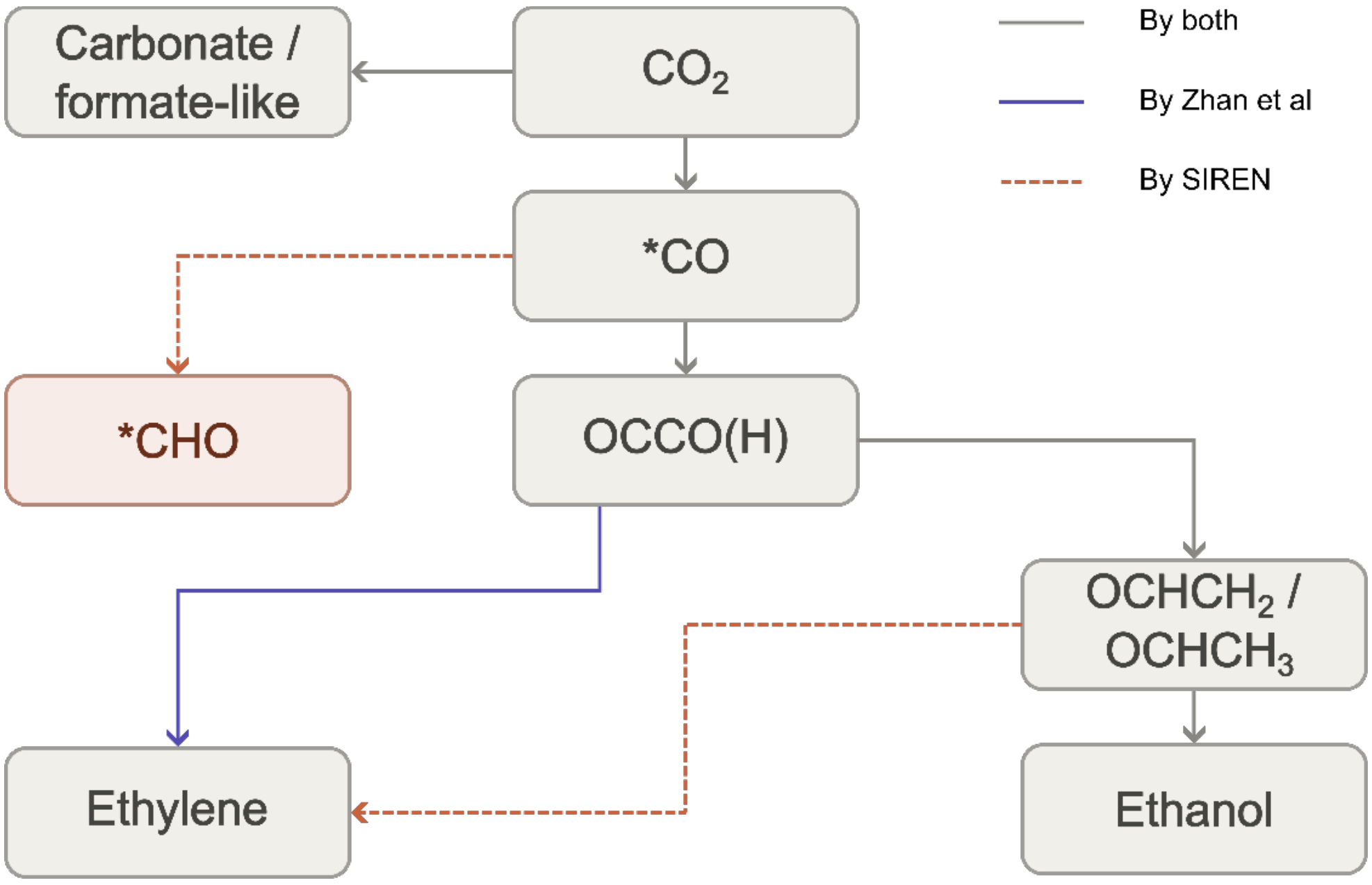

**Figure S10**. Schematic illustration of $CO_2$ electroreduction pathways. Gray arrows indicate general reaction pathways proposed by both, the blue arrow highlights the OCCO(H)-derived ethylene pathway suggested by Zhan et al., and orange dashed arrows represent alternative pathways suggested by the SIREN.

**Table S15**. Performance of SIREN with additional language model in the computational benchmarks.

| | System | Average Best Spectrum Distance | Success Rate |
|---|---|---|---|
| SIREN (GPT-5-mini) | Cu (111) | 0.016 | 0.7 |
| | Cu3Pd (111) | 0.045 | 0.4 |
| | Fe3Cu (111) | 0.075 | 0.0 |
| | Cu (211) | 0.027 | 0.6 |
| | All | 0.041 | 0.425 |
| SIREN (GPT-6-luna) | Cu (111) | 0.019 | 0.7 |
| | Cu3Pd (111) | 0.058 | 0.1 |
| | Fe3Cu (111) | 0.070 | 0.0 |
| | Cu (211) | 0.027 | 0.5 |
| | All | 0.044 | 0.325 |

**Table S16**. Potential-dependent adsorbate compositions of final surface states selected in independent SERS-guided inference runs on OD Cu.

| Model | Potential ($V_{RHE}$) | Inferred Adsorbate Species |
| --- | --- | --- |
| gpt-5-mini | −0.36 | [OCCHO, HCOO], [$OCHCH_3$, HCOO], [$CO_3$, CH, COOH], [HCOOH, OCCHO], [$OCHCH_3$, $OCHCH_2$] |
| | −0.56 | [$OCHCH_2$, CHO], [OCCHO, $OCHCH_3$], [HCOO, $OCHCH_3$], [OCCHO, HCOO], [OCCHO, CH, $OCHCH_3$] |
| | −0.76 | [OCCHOH, $OCHCH_2$], [CO, CH, $OCHCH_2$], [CO, OCCHO, OCCO], [CO, 2 × HCOO], [CO, CH, HCOO] |
| | −0.85 | [CO, $OCHCH_3$], [CO, OCCHO], [CO, $OCHCH_3$], [CO, HCOO], [CO, $CH_3$, HCOO] |
| | −0.95 | [CO, $OCHCH_3$], [CO, $OCHCH_2$], [CO, $OCHCH_3$], [CO, CH, HCOO], [CO, $OCHCH_3$] |
| gpt-5.6-luna | −0.36 | [2 × $CO_3$, HCOO], [$CO_3$, HCOO], [$CO_3$, HCOO], [$CO_3$, CH, HCOO], [CHO, $OCHCH_2$] |
| | −0.56 | [$CO_3$, CHO, HCOOH], [$CO_3$, OCCO, $OCHCH_3$], [OCCHO, $OCHCH_2$, CO], [OCCO, $OCHCH_3$], [$CO_3$, OCCHO] |
| | −0.76 | [CO, CHO, HCOO], [CO, HCOO, OCCO], [CO, HCOO, $OCHCH_3$], [CO, $CO_3$, HCOOH, $OCHCH_2$], [CO, $OCHCH_2$, OCCHO] |
| | −0.85 | [CO, CHO, HCOO, $OCHCH_3$], [HCOO, OCCOH], [$CO_3$, HCOO, OCCO], [CHO, HCOO, OCCHOH, $OCHCH_3$], [CO, CHO, HCOO] |
| | −0.95 | [CO, CH, HCOO, OCCHO], [CHO, OCCOH, $OCHCH_2$], [CO, HCOO, $OCHCH_2$], [OCCHOH, HCOO, $OCHCH_3$], [$CO_3$, OCCHOH, CH] |
| gpt-6-luna | −0.36 | [$OCHCH_3$], [$CO_3$, $OCHCH_3$], [$CO_3$, OHCCOH], [$CO_3$, HCOOH], [$CO_3$, $OCHCH_3$] |
| | −0.56 | [$CO_3$, HCOO, CO], [CO, $CO_3$, OHCCOH] [HCOO, OCCO, $CO_3$], [$CO_3$, $OCHCH_3$], [$CO_3$, HCOOH, CO] |
| | −0.76 | [HCOO, OCCHOH], [$CO_3$, HCOO, OCCO], [OCCHOH, $CO_3$, HCOO], [OHCCOH, OCCHOH, HCOO], [OCCHOH, CH] |
| | −0.85 | [HCOO, OCCHOH, CH], [OCCHOH, $CO_3$, HCOO, CH], [OCCHOH, HCOO, CH], [OCCHOH, $OCHCH_2$, $CO_3$, CH], [OCCHOH, $OCHCH_2$] |
| | −0.95 | [CO, CH, $OCHCH_2$], [OCCHOH, HCOO], |
| | | [OCCO, CHO, HCOO], [OCCHOH, $OCHCH_2$], [OCCHOH, HCOOH, CH] |

**Table S17**. Raw responses about mechanistic interpretation of LLMs.

| Model | Response |
|---|---|
| gpt-5-mini | Summary of approach and caveats<br>- I treat the AI-inferred labels (OCCHO, OCHCH3, OCCO, etc.) as tentative spectroscopic assignments, not definitive chemical identities or coverages. I rely on chemically plausible interpretations (OCCHO = oxygenated C2 acyl/aldol-like adduct; OCCO = *CO–*CO dimer; OCHCH3/OCHCH2 = alkoxy/vinylic C2 fragments) and emphasize where the spectrum-derived picture supports, contradicts, or is ambiguous relative to the measured Faradaic efficiencies (FE).<br>- The FE figure and the inferred surface states come from different experiments; I therefore treat FE as contextual evidence and avoid forcing agreement—discrepancies may reflect differences in conditions, spectral selection rules, or site heterogeneity.<br><br>A. Potential-by-potential comparison of surface states and FE trends<br><br>Notes on the FE figure (visual read)<br>- Potentials shown (left → right): about −1.10, −1.05, −1.00, −0.95, −0.90, −0.80 V (approx.; read visually).<br>- General trend: at the most negative potentials (≈ −1.00 to −1.10 V) C2H4 is the single largest product; CH4 and EtOH are also significant. At less negative potentials (≈ −0.90 to −0.80 V) H2 and formate are large contributors and CO shows a larger share than at the most negative potentials.<br>- Selected approximations (visual): at ≈ −1.00 V C2H4 ≈ 30–45%, CH4 ≈ 15–30%, EtOH ≈ 5–12%; at ≈ −0.90 V H2 ≈ 25–35%, formate ≈ 20–35%, CO ≈ 10–20%. (I report ranges; I do not force sums to 100%.)<br><br>Potential alignment (which potentials have both data)<br>- Surface-state potentials available: −0.36, −0.56, −0.76, −0.85, −0.95, −1.12 V vs RHE.<br>- FE figure potentials (visual): ≈ −1.10, −1.05, −1.00, −0.95, −0.90, −0.80 V.<br>- Overlaps / near-overlaps (qualitative):<br>- Surface −1.12 V ≈ FE −1.10 V → comparable.<br>- Surface −0.95 V ≈ FE −0.95 V → comparable.<br>- Surface −0.85 V is between FE −0.90/−0.80 → compare qualitatively with those FE points (do not claim exact match).<br>- Surface −0.76, −0.56, −0.36 V are outside the FE potential window shown; comparisons to FE at −0.80/−0.90 are therefore qualitative only.<br><br>Potential-by-potential notes<br><br>- −0.36 V (surface data only)<br>- Surface inference: formate-related species (HCOO / HCOOH) most reproducible; some oxygenated C2 (OCCHO, OCHCH3) weakly present. CO and OCCO absent.<br>- FE: no FE point at −0.36 V in the figure. Contextual note: formate appearance in the inferred states at mild overpotential is chemically reasonable, but cannot be directly tied to the FE bars shown.<br><br>- −0.56 V (surface data only)<br>- Surface: increasing OCCHO and OCHCH3 assignments; formate still present but less dominant.<br>- FE: no direct FE at −0.56 V. This inferred shift toward oxygenated C2 intermediates is not contradicted by FE (FE at more negative potentials shows both oxygenates and C2 hydrocarbons).<br><br>- −0.76 V (surface data only; closest FE point is −0.80/−0.90)<br>- Surface: strong rise in CO signal, first consistent OCCO (*CO–*CO) inference; OCCHO/OCCHOH, OCHCH2 also present; CH/CHO begin to appear.<br>- FE (nearby at −0.80 to −0.90): FE in this window shows substantial formate + H2 (at −0.90) but moving more negative toward −1.00 the dominant product shifts to C2H4. The surface CO/OCCO appearance at −0.76 is consistent with enabling C2 formation seen in FE at more negative potentials, but direct quantitative mapping is not established.<br><br>- −0.85 V (surface) ≈ compare to FE −0.90/−0.80 |

- Surface: CO dominant in most runs; oxygenated C2 fragments (OCHCH3/OCCHO) persist.
- FE at −0.90: H2 and formate are large, CO non-negligible, C2H4 moderate. Apparent mismatch: surface shows CO dominance at −0.85, while FE at nearby −0.90 shows larger FE to H2/formate and only moderate C2; this is not necessarily contradictory because surface CO may be reactive (converted to C2 products) and not appear as CO FE, and spectral features can be biased toward species with strong Raman cross-sections.

- −0.95 V (surface) ≈ FE −0.95 V
  - Surface: CO very reproducible; oxygenated C2 fragments (OCHCH3) reproducible; some HCOO still inferred; small CH signal.
  - FE at −0.95: approaching the region where C2H4 and EtOH/CH4 are substantial (C2 increases going more negative). The inferred presence of *CO and oxygenated C2 intermediates supports the CO-mediated C–C coupling route and oxygen-retaining C2 pathways that can produce ethanol/ethylene.

- −1.12 V (surface) ≈ FE −1.10 V
  - Surface: CO still present; OCCO and OCCHO appear in some runs; CH/CHO and multi-O species (OHCCOH) appear—sign of deeper hydrogenation and complex oxygenated intermediates.
  - FE at ≈ −1.10 V: C2H4 is the dominant FE at the most negative potentials in the figure; CH4 and EtOH also appear. The surface emergence of hydrogenated C1/C2 fragments (CHO/CH) is consistent with increased CH4 formation and with intense C–C coupling at these potentials.

B. Comparison table (condensed)

| Potential | Surface-state evidence (inferred) | Major FE products (visual) | Mechanistic interpretation | Agreement / discrepancy | Limitations |
|---|---|---|---|---|---|
| −0.36 V | HCOO/HCOOH (most reproducible); OCCHO/OCHCH3 observed | (no FE shown) | Formate pathway active at mild overpotential; early oxygenated C2 fragments possible | N/A (no FE at same potential) | Different potential ranges; cannot validate with FE figure |
| −0.56 V | OCCHO, OCHCH3 rising; HCOO present | (no FE shown) | Shift toward oxygenated C2 chemistry begins | N/A | As above |
| −0.76 V | CO increases; OCCO first appears; OCCHO/OCCHOH present; CH/CHO emergent | FE nearby (−0.80/−0.90): H2 & formate relatively large; C2 moderate | CO accumulation can enable *CO–*CO coupling (OCCO) → C2; early hydrogenation branches appear | Partial agreement: OCCO/CO consistent with later C2 increase; FE at −0.90 still formate/H2-dominant—could reflect kinetics/site heterogeneity | Surface vs FE measured at nearby but not identical potentials; spectral vs product sensitivities differ |
| −0.85 V | CO dominant; oxygenated C2 fragments persist | FE at ≈−0.90: H2 & formate large; CO moderate; C2 modest | CO-rich surface can be reactive intermediate reservoir feeding C2 formation (not necessarily high CO FE) | Apparent mismatch (surface CO high vs FE CO moderate) is plausible if CO is converted to C2 products or if CO has strong spectral signature | Spectroscopic cross-section bias; transient kinetics not captured |
| −0.95 V | CO high; OCHCH3 common; HCOO still seen; small CH | FE at −0.95: rising C2H4 & EtOH; CH4 present | Coexistence of *CO, OCCO-like and oxygenated C2 species supports both CO-dimerization and oxygen-retaining routes to ethanol/ethylene | Agreement: intermediates required for C2 formation are inferred where FE shows significant C2 | Inferred species are tentative; branching not resolved by spectrum alone |
| −1.12 V | CO present; OCCO & OCCHO appear in some runs; CH/CHO and multi-O species appear | FE at −1.10: C2H4 dominant; CH4 & EtOH significant | More negative potential increases hydrogenation steps (toward CH4) and promotes C–C coupling → C2 products; oxygenated intermediates may be further hydrogenated | Agreement: hydrogenation signatures (CHO/CH) align with CH4 FE; C2 intermediates align with large C2 FE | Single-scan spectra cannot quantify flux; alternative assignments (e.g., spectators) possible |

C. Potential-dependent pathway interpretation — supported, plausible, speculative

Supported by combined evidence (spectrum + FE)
- CO-mediated C–C coupling: Appearance of strong CO signals and OCCO at intermediate/negative

potentials (−0.76 to −1.12) together with high $C_2H_4$ FE at ≈ −1.00 to −1.10 supports that *CO accumulation and *CO–*CO coupling are important routes to C2 products on oxide-derived Cu.
- Formate pathway at mild potentials: Reproducible HCOO/HCOOH spectral signatures at mild overpotential (−0.36) and significant FE to formate at less-negative FE points (≈ −0.90 to −0.80) support the formate branch under low overpotential or certain surface sites.

Plausible (chemically consistent, but not uniquely proven)
- Oxygen-retaining C–C coupling (OCCHO/OCCHOH → OCHCH3/OCHCH2 → ethanol/acetate): OCCHO/OCHCH3 inferred across potentials and presence of ethanol FE at ≈ −1.00 suggest oxygenated C2 intermediates exist and can lead to oxygenate products depending on hydrogenation/C–O cleavage balance.
- Competition between hydrogenation (→ $CH_4$) and deoxygenation (→ $C_2H_4$) of C2 adducts: increasing CH/CHO features at −1.12 align with greater $CH_4$ FE at most negative potentials; plausible branching controlled by local H activity.

Speculative (require targeted validation)
- Specific sequences (e.g., OCCO → OCCHO → OCHCH3 as the dominant chemical transformation) and assignments of multi-oxygen species like “OHCCOH” as on-path intermediates are speculative without vibrational assignment validation and DFT-calculated frequencies.
- Exact site assignments (which facet or subsurface oxygen sites host the OCCHO vs OCCO pathway) require surface-specific probes.

D. Three most important mechanistic insights (each with observation, interpretation, alternative, evidence needed)

1) Insight 1 — Transition from formate to CO-mediated C2 chemistry with increasing overpotential
- Observation: Spectra show HCOO/HCOOH most reproducible at −0.36 V; CO and OCCO signals become reproducible from −0.76 V and above. FE shows high formate/$H_2$ contributions at less negative potentials (≈ −0.90–−0.80) and dominant $C_2H_4$ at most negative potentials (≈ −1.00 to −1.10).
- Mechanistic interpretation: At low overpotential formate pathway is competitive or dominant; increasing overpotential increases *CO formation/coverage, enabling *CO–*CO coupling and C2 formation.
- Alternative explanation: Apparent formate spectral intensity at mild potentials might reflect strongly Raman-active spectator formate/carbonate bound near the surface, rather than a kinetically important pathway; similarly, surface CO at negative potentials could reflect slow desorption rather than increased formation rate.
- Evidence needed: Time-resolved potential-step experiments (operando SERS/IR + simultaneous product detection) and $^{13}CO_2$ labeling to show conversion flux from surface CO/OCCO to C2 products.

2) Insight 2 — Coexistence of oxygen-retaining C2 intermediates with CO-dimer pathways; branching controls ethanol vs ethylene
- Observation: OCCHO / OCCHOH and alkoxy-like OCHCH3/OCHCH2 assignments recur across mid-to-negative potentials; FE shows ethanol present together with large $C_2H_4$ FE at similar potentials.
- Mechanistic interpretation: Oxide-derived Cu stabilizes oxygenated C2 adducts that can be hydrogenated to oxygenates (ethanol/acetate) or undergo C–O cleavage to yield hydrocarbons (ethylene). Both routes may coexist and compete.
- Alternative explanation: The oxygenated spectral features could be spectator species or products readsorbed on the surface (e.g., adsorbed acetate/alkoxide) rather than true on-path intermediates.
- Evidence needed: Isotope labeling ($^{13}C$, $^{18}O$) in combination with operando vibrational spectroscopy to track oxygen retention/loss, and DFT vibrational assignments to validate spectral IDs.

3) Insight 3 — Increasing hydrogenation at most negative potentials drives $CH_x$ signatures and $CH_4$ formation
- Observation: CH and CHO-like inferred species and more reduced multi-O fragments appear at −1.12 V; FE at −1.10 shows substantial $CH_4$ in addition to C2.
- Mechanistic interpretation: Higher overpotentials increase hydrogenation rates relative to C–O bond scission, promoting C1 hydrogenation to methane and deeper hydrogenation of C2 intermediates; this

| | |
|---|---|
| | shifts some flux away from certain oxygenate or deoxygenation pathways.<br>- Alternative explanation: The inferred CH/CHO signals could be localized species at defect sites not representative of bulk activity, or the spectral assignments could be ambiguous.<br>- Evidence needed: Potential-step transients correlating the rise of CH/CHO spectral signals with increases in CH4 formation; kinetic isotope effect using D2 to test hydrogenation-limited steps.<br><br>E. 150–200 word Results & Discussion paragraph (suitable for a manuscript)<br><br>Operando spectral inference and independent product analysis together indicate a potential-dependent mechanistic bifurcation on oxide-derived Cu. At mild overpotentials the surface spectrum is dominated by formate-related signatures, consistent with the high formate/hydrogen FE observed at less negative potentials. As the potential is made more negative, *CO becomes the dominant spectroscopic species and OCCO-like signatures appear; this correlates with the emergence of high C2H4 FE at ≈ −1.00 to −1.10 V, supporting a CO-mediated dimerization route toward C2 products. Persistently inferred oxygenated C2 fragments (OCCHO, OCHCH3/OCHCH2) across several potentials provide independent evidence that oxygen-retaining C–C adducts coexist with pure *CO–*CO chemistry and plausibly branch toward ethanol or deoxygenated C2 products depending on hydrogenation and C–O cleavage rates. At the most negative potentials increased CH/CHO spectral features and multi-oxygen fragments are consistent with enhanced hydrogenation, which can divert flux toward CH4 and heavily hydrogenated oxygenates. Crucially, spectral detection of an intermediate does not equate to high FE; discrimination between spectator versus on-path species and between competing branches requires time-resolved isotopic labeling and coordinated product detection.<br><br>F. Recommended targeted experiments (brief)<br>- Rapid potential-step operando SERS/IR coupled to fast GC/DEMS to order intermediates temporally.<br>- 13CO2 (and 18O) labeling with operando vibrational spectroscopy to test oxygen retention in C2 products.<br>- DFT-calculated vibrational spectra for candidate OCCHO/OCCO/OCHCHx species and energetics for branching steps (hydrogenation vs C–O cleavage).<br>- Facet-controlled or site-blocking studies to test whether formate vs CO–dimer pathways are site-specific.<br><br>Final caution<br>- The surface-state assignments are tentative and spectroscopically biased; FE trends are from a different study and must be treated as contextual. The comparisons above highlight chemically plausible consistencies and concrete discrepancies; resolving them requires the targeted experiments noted. |
| gpt-5.6-luna | ## A. Potential-by-potential comparison<br><br>### Reading of the FE figure<br>The plotted FE values are approximate because several segment boundaries are close to the graph resolution. The dominant trends are:<br><br>- **−1.10 to −1.00 V:** $C_2H_4$ is the largest $CO_2$RR product, approximately 34–47%; $CH_4$ is also substantial, roughly 20–25%. $H_2$ is about 10–26%. Ethanol is present at lower FE, approximately 5–10%.<br>- **−0.95 V:** $C_2H_4$ remains prominent at approximately 40%, while $CH_4$ decreases to roughly 10–15%. $H_2$ is approximately 15%, and CO, formate, ethanol, and 1-propanol are minor-to-moderate products.<br>- **−0.90 V:** The product distribution shifts strongly toward $H_2$, CO, and formate. Approximate values are $H_2$ 30–33%, formate 24–26%, CO 15–17%, and $C_2H_4$ approximately 14–16%. Ethanol is small.<br>- **−0.80 V:** $H_2$ and formate/CO dominate, approximately 35–40%, 25–30%, and 20–22%, respectively. $C_2H_4$ is only a few percent, and ethanol and methane are not prominent.<br><br>The FE and surface-state potentials overlap exactly only at **−0.95 V**. The remaining comparisons are between neighboring but nonidentical potentials.<br><br>### −0.36 V vs RHE — surface-state data only<br><br>The inferred state is dominated by `CO3` and `HCOO`, with `CO3` present in all five runs and `HCOO` in |

four. The isolated `CH`, `CHO`, and `OCHCH2` assignments provide weak evidence for reduced carbon chemistry, but not for a substantial $C_2$ pathway.

No FE point is available near this potential in the figure. Therefore, it is not possible to determine whether the formate-like surface chemistry corresponds to measurable formate production, spectator bicarbonate/formate equilibria, or a low-flux adsorbed reservoir. The isolated reduced-carbon assignments should not be used to infer early $C_2$ production.

### −0.56 V vs RHE — surface-state data only

`OCCO`, `OCCHO`, and `OCHCH3` recur in two of five runs each. This is the strongest direct spectral indication of a possible C–C-coupled oxygenated surface ensemble. A pathway involving CO-derived coupling is chemically plausible, but the structures are tentative and their relative sequence is unknown.

There is no FE point at this potential. Consequently, the inferred $C_2$-related states cannot be directly linked to ethanol, ethylene, or any other product. They may be reactive intermediates, stable spectators, minority sites selectively enhanced by SERS, or species formed under conditions different from those used for the FE measurements.

### −0.76 V vs RHE — surface-state data; neighboring FE at −0.80 V

`CO` is assigned in all five runs, while `HCOO` occurs in three. This supports a persistent CO-like surface pool and a parallel formate-like branch. `OCCO` and `OCHCH2` occur only occasionally.

The neighboring FE point at −0.80 V is dominated by $H_2$, formate, and CO, with only a few percent $C_2H_4$. This is consistent with a substantial CO reservoir that does not necessarily proceed efficiently to $C_2$ products. It may reflect CO accumulation, slow C–C coupling, competitive desorption, or a change in surface state between −0.76 and −0.80 V. The apparent CO pool therefore supports CO involvement but not high $C_2$ flux.

### −0.85 V vs RHE — surface-state data; neighboring FE at −0.90 V

`HCOO` is present in every run and `CHO` in three. `CO` and `OCHCH3` occur in two runs. The data support concurrent formate-like and hydrogenated carbonyl chemistry, with a tentative oxygenated $C_2$ branch.

At the neighboring −0.90 V FE point, formate and CO are substantial, $C_2H_4$ is still measurable at approximately 15%, and ethanol is minor. This partially agrees with the inferred coexistence of formate and reduced CO-derived species. However, the relatively low ethanol FE does not establish that `OCHCH3` is an ethanol precursor. The same assignment could represent a surface configuration on a deoxygenating or nonproductive branch.

### −0.95 V vs RHE — exact surface-state and FE overlap

This is the most informative comparison because both datasets are available at the same nominal potential. The inferred surface contains `CO`, `HCOO`, and `OCHCH2` in two of three runs, with isolated `OCCHO` and `OCCOH`.

The FE data show high $C_2H_4$ production, approximately 40%, alongside substantial $CH_4$, $H_2$, and smaller amounts of formate, CO, ethanol, and 1-propanol. The coexistence of `OCHCH2` with high $C_2H_4$ is compatible with a hydrogenated $C_2$ intermediate pool that can undergo deoxygenation toward ethylene. However, this is not a unique assignment: `OCHCH2` could also be an oxygen-retaining $C_2$ intermediate, and the modest ethanol FE indicates that oxygen-retaining chemistry is present but not dominant under these conditions.

The inferred formate does not correspond to especially high formate FE at this potential, emphasizing that detected surface species need not be the principal product-forming flux.

### −1.12 V vs RHE — surface-state data; neighboring FE at −1.10 V

`CHO`, `HCOO`, and `OCCOH` are the most reproducible assignments. This is consistent with increased hydrogenation of CO-derived species and persistence of an oxygenated $C_2$ branch. The isolated `CO` and `OCCO` assignments do not support a large, spectroscopically persistent free-CO or coupling precursor pool at this potential.

At −1.10 V, the FE distribution is dominated by $C_2H_4$ and $CH_4$, with substantial $H_2$ and lower ethanol FE. The presence of `CHO` is compatible with the observed $CH_4$ branch, although `CHO` is not methane-specific. `OCCOH` is compatible with $C_2$ chemistry, but cannot distinguish ethylene from ethanol formation. The high $C_2H_4$ FE despite the low recurrence of `OCCO` suggests that C–C coupling intermediates may be rapidly consumed, spectroscopically weak, or represented by assignments not captured in the inferred states.

### FE-only potentials

- **−1.05 V:** $C_2H_4$ remains dominant, approximately 40%, with substantial $CH_4$ and lower $H_2$. No directly inferred surface state is available.
- **−1.00 V:** $C_2H_4$ is highest, approximately 45–47%, with $CH_4$ around 20–25%, moderate ethanol, and lower $H_2$. No directly inferred surface state is available.
- **−0.90 V:** $H_2$, formate, CO, and $C_2H_4$ are all significant. The nearby −0.85 V surface data suggest plausible correspondence to formate and hydrogenated carbon chemistry, but no direct equality of potentials or conditions should be assumed.
- **−0.80 V:** $H_2$, formate, and CO dominate, while $C_2H_4$ is minor. The nearby −0.76 V CO-rich inferred state is compatible with this distribution but does not establish that adsorbed CO is the source of the measured CO product.

---

## B. Comparison table

| Potential | Surface-state evidence | Major FE products | Mechanistic interpretation | Agreement/discrepancy | Limitations |
|---|---|---|---|---|---|
| −0.36 V | `CO3` 5/5; `HCOO` 4/5; isolated `CHO`, `OCHCH2` | No FE point | Carbonate/formate-like interfacial chemistry; weak evidence for reduced carbon | Cannot be tested against FE | Carbonate may be spectator; formate frequency is not coverage or flux |
| −0.56 V | Repeated `OCCO`, `OCCHO`, `OCHCH3` | No FE point | Plausible CO-derived C–C coupling and oxygenated $C_2$ network | Mechanistically suggestive but unvalidated | Species are tentative and not product-specific |
| −0.76 V | `CO` 5/5; `HCOO` 3/5; occasional `OCCO`, `OCHCH2` | At −0.80 V: $H_2$, formate, CO dominant; $C_2H_4$ minor | Persistent CO pool with competing formate and limited $C_2$ flux | CO-rich surface is consistent with CO FE, but not with high ethylene FE | Neighboring rather than identical potential; adsorbed CO may accumulate |
| −0.85 V | `HCOO` 5/5; `CHO` 3/5; `CO`, `OCHCH3` 2/5 | At −0.90 V: $H_2$/formate dominant, CO and $C_2H_4$ moderate | Parallel formate, CO/CHO hydrogenation, and possible $C_2$ oxygenate chemistry | Broad agreement with mixed FE distribution; ethanol remains low | Spectral labels do not identify ethanol or ethylene pathways |
| −0.95 V | `CO`, `HCOO`, `OCHCH2` in 2/3; isolated `OCCHO`, `OCCOH` | $C_2H_4$ approximately 40%; $CH_4$ and $H_2$ also substantial | Hydrogenated $C_2$ pool is compatible with strong ethylene formation; parallel C1 and formate chemistry | Strongest alignment between inferred $C_2$ species and a $C_2$ product | Only three runs; `OCHCH2` is non-unique; FE publication may differ in conditions |
| −1.05 V | No surface data | $C_2H_4$ dominant, roughly 40%; $CH_4$ also high | $C_2$ and C1 reduction both favorable | Cannot directly compare | No operando surface assignment |
| −1.00 V | No surface data | $C_2H_4$ dominant, roughly 45–47%; $CH_4$ substantial | High $C_2$ hydrocarbon selectivity | Mechanism cannot be assigned from supplied spectra | Missing surface information |

| −0.90 V | No exact surface data; neighboring −0.85 V state | $H_2$, formate, CO, $C_2H_4$ significant | Competition among HER, CO formation, formate chemistry, and $C_2$ reduction | Qualitatively compatible with nearby surface assignments | Potential and experimental conditions may differ |
| −0.80 V | No exact surface data; neighboring −0.76 V state | $H_2$, formate, CO dominant; $C_2H_4$ minor | CO retention/desorption and formate/HER competition outweigh $C_2$ production | Consistent with a persistent CO pool but does not prove its catalytic role | No direct surface measurement at −0.80 V |
| −1.10 V | No exact surface data; neighboring −1.12 V state | $C_2H_4$ and $CH_4$ dominant; $H_2$ substantial; ethanol lower | Hydrogenated C1 and C2 branches operate at high negative bias | `CHO`/`OCCOH` assignments are broadly compatible | Neighboring potential; no quantitative relation between recurrence and FE |

---

## C. Potential-dependent pathway interpretation

### More directly supported

1. **A formate/carbonate-like interfacial population exists over a broad potential range.**
The repeated `CO3` and `HCOO` assignments support persistent interfacial or adsorbed oxygen-containing species. Their exact catalytic role is unresolved.

2. **A CO-like surface pool is particularly reproducible near −0.76 V.**
This is consistent with the measurable CO FE near −0.80 V, but it does not establish whether CO is a productive intermediate, a product that remains adsorbed, or a spectroscopically favored reservoir.

3. **$C_1$ and $C_2$ reduction pathways operate concurrently at negative potentials.**
The FE data show substantial $CH_4$ and $C_2H_4$ simultaneously, while the spectra contain `CHO` and multiple tentative $C_2$ assignments.

### Plausible but not demonstrated

- CO-derived C–C coupling through an `OCCO`-like species.
- Hydrogenation of CO-derived species toward `CHO`.
- Formation of hydrogenated $C_2$ intermediates represented by `OCHCH2`, `OCHCH3`, `OCCOH`, or `OCCHO`.
- Competition between oxygen-retaining $C_2$ chemistry and deoxygenation toward $C_2H_4$.

### Speculative connections

- A specific sequence `CO → OCCO → OCCHO/OCCOH → OCHCH2/OCHCH3`.
- Assignment of `OCHCH2` specifically to an ethylene precursor or `OCHCH3` specifically to ethanol.
- Direct conversion of the inferred `HCOO` population into measured formate.
- A causal relationship between the −0.56 V $C_2$ assignments and the FE maxima at −1.00 to −0.95 V, because no FE or surface-state data are available at matching conditions.

---

## D. Three most important mechanistic insights

### 1. High $C_2H_4$ FE is compatible with, but not proven by, the inferred $C_2$ oxygenated species

- **Observation:** At the exact overlap of −0.95 V, `OCHCH2` is repeated in two of three runs, while $C_2H_4$ is the dominant FE product at approximately 40%.
- **Mechanistic interpretation:** A hydrogenated $C_2$ surface pool may undergo deoxygenation to ethylene. This provides a chemically plausible connection between the spectral and product data.
- **Alternative explanation:** `OCHCH2` may be an oxygen-retaining intermediate, a nonproductive adsorbate, or a misassigned vibrational structure. $C_2H_4$ may form through a different intermediate that is not observed.
- **Evidence needed:** Isotopically labeled $CO_2$/CO, potential-dependent isotope-resolved vibrational spectra, independent vibrational calculations for competing $C_2$ structures, and correlation of transient intermediate signals with ethylene evolution.

### 2. Persistent adsorbed CO does not imply high $C_2$ flux

- **Observation:** `CO` is assigned in all runs at −0.76 V, whereas neighboring −0.80 V conditions show high CO and $H_2$ FE but only a few percent $C_2H_4$.
- **Mechanistic interpretation:** CO may be a necessary precursor but be kinetically trapped, desorb before coupling, or coexist with strong HER and formate pathways.
- **Alternative explanation:** The spectral assignment may preferentially detect CO at SERS-active sites, while the productive C–C coupling sites are rare or spectroscopically invisible. The neighboring potentials may also correspond to different reconstructed surfaces.
- **Evidence needed:** Operando CO coverage measurements, CO stripping or perturbation experiments, controlled CO-feed experiments, and product transients following $CO_2$-to-CO switching.

### 3. The CHO/OCCOH-rich negative-potential state is consistent with mixed C1/C2 chemistry, not a unique product pathway

- **Observation:** At −1.12 V, `CHO` and `OCCOH` are frequent, while the neighboring −1.10 V FE contains substantial $CH_4$ and $C_2H_4$, with lower ethanol.
- **Mechanistic interpretation:** `CHO` may participate in deep C1 hydrogenation toward methane, while `OCCOH` may represent an oxygenated $C_2$ branch that partly deoxygenates to ethylene.
- **Alternative explanation:** These assignments could reflect protonation-state changes, altered SERS enhancement, or stable off-cycle species. CHO is not methane-specific, and OCCOH is not ethanol- or ethylene-specific.
- **Evidence needed:** Product-resolved isotope tracing, hydrogen/deuterium substitution, pH and cation variation, vibrational Stark analysis, and DFT-assisted discrimination of C1 versus C2 adsorption geometries.

---

## E. Scientific discussion

The spectrum-derived surface states provide qualitative support for parallel formate-like, CO-mediated, C1 hydrogenation, and C2 pathways, but they do not uniquely determine product flux. Formate-like assignments are frequent over much of the potential range, whereas a persistent CO assignment is observed near −0.76 V. The neighboring FE data at −0.80 V are dominated by $H_2$, formate, and CO, with only minor ethylene, indicating that a detectable CO reservoir does not necessarily correspond to efficient C–C coupling. The strongest correspondence occurs at −0.95 V, where repeated `OCHCH2` assignments coincide with approximately 40% $C_2H_4$ FE. This is consistent with a hydrogenated $C_2$ population undergoing deoxygenation, although the assignment is not unique and cannot exclude oxygen-retaining pathways. At more negative potential, frequent `CHO` and `OCCOH` assignments are compatible with simultaneous methane and ethylene production observed near −1.10 V, but neither species is product-specific. The absence of a reproducible `OCCO` signal at these potentials may reflect rapid consumption, low coverage, SERS selection, or altered surface structure rather than the absence of C–C coupling. Because the FE data originate from another publication, differences in catalyst morphology, electrolyte, calibration, and reconstruction prevent direct quantitative validation.

gpt-6-luna

Below I compare the AI-inferred surface states (treated as tentative spectral assignments) to the Faradaic efficiency (FE) bar chart in the figure, following your requested structure and the required caveats. I treat the inferred species names as motifs (CO*, HCOO*, OCCO*, OCCHO-like, OCHCHx, CH/CHO, CO3/HCO3) and distinguish evidence-supported statements from hypotheses and speculation.

1) Experimental FE interpretation (visual reading of the figure)
- Potentials plotted (approx.): −1.10, −1.05, −1.00, −0.95, −0.90, −0.80 V (labelled on x axis).
- Dominant products and qualitative FE trends (visually distinguishable):

- Most negative potentials (−1.10 to −1.00 V): large $C_2H_4$ FE (dark blue) is a major product. $CH_4$ (light green) and $H_2$ (brown) are also significant; ethanol (EtOH, orange) and small 1-propanol (light blue) appear at measurable but lower FE.
- Around −0.95 V: $C_2H_4$ remains large; $H_2$ and $CH_4$ are present at moderate FE; EtOH present at moderate-low FE.
- At less negative potentials (−0.90 to −0.80 V): formate (green) and $H_2$ (brown) increase in FE; CO (yellow) and $C_2H_4$ decline relative to the most negative potentials. At −0.80 V, formate and $H_2$ are among the largest single contributions.

- I report these as qualitative/relative observations; I do not force totals to 100% or invent precise numbers.

2) Potential alignment (which data overlap)

- Potentials with both surface-state and FE data (close enough to compare qualitatively):
  - Surface −1.12 V ↔ FE −1.10 V (neighboring; compare qualitatively).
  - Surface −0.95 V ↔ FE −0.95 V (direct match).
- Surface-only potentials (no corresponding FE bar at the same labeled potential): −0.36 V, −0.56 V, −0.76 V, −0.85 V. These can be compared only to nearby FE points (−0.80 and −0.90 V) qualitatively.
- FE-only potentials (no inferred surface states at those exact potentials): −1.05 V, −1.00 V, −0.90 V, −0.80 V (FE data exist at these potentials but there are no separate AI surface reconstructions listed for them).
- When potentials are not identical I compare qualitatively (e.g., surface at −0.76 V with FE at −0.80 or −0.90 V) and explicitly note the mismatch.

A. Potential-by-potential analysis (surface chemistry vs FE)

- Surface −1.12 V (compare to FE at −1.10 V)
  - Surface evidence: CO*, OCCO-type (CO–CO coupled) motif, OCCHO/OCCHOH, CH/CHO, occasional multi-O C2 motif (OHCCOH). CO inferred frequently.
  - FE (−1.10 V): very large $C_2H_4$ FE, substantial $CH_4$ and $H_2$ FE; measurable EtOH and small longer alcohols.
  - Mechanistic alignment: Presence of CO* and OCCO supports an on-surface CO-dimerization route toward C2 products (consistent with high $C_2H_4$ FE). CH/CHO on the surface is consistent with observed $CH_4$ FE. Coexistence of oxygenated C2 motifs (OCCHO etc.) is consistent with some EtOH FE. Agreement is good qualitatively.
  - Caveats: OCCO inferred in only some runs (spectroscopic intermittency), and detection does not quantify flux to ethylene vs ethanol.

- Surface −0.95 V (direct match to FE −0.95 V)
  - Surface evidence: CO* dominant, $OCHCH_3$ / $OCHCH_2$ (hydrogenated oxygenated C2 motifs), HCOO present in some runs, occasional CH.
  - FE (−0.95 V): $C_2H_4$ still a major product (though the exact magnitude shifts across −1.00 to −0.95), EtOH present at lower but non-zero FE; $CH_4$ and $H_2$ present.
  - Mechanistic alignment: CO* dominance and detected C2 oxygenated fragments ($OCHCH_3$) are consistent with active CO-mediated C–C coupling and partial hydrogenation toward oxygenates (explaining EtOH). The spectroscopic detection of O-containing C2 motifs and CO* is compatible with mixed outcomes (ethylene > ethanol).
  - Caveats: $OCHCH_3$ motif is ambiguous — it could be an on-path hydrogenated intermediate (→ ethanol) or a surface-stabilized oxygenate that does not desorb efficiently.

- Surface −0.85 V and −0.76 V (compare to FE −0.80 to −0.90 V)
  - Surface evidence at −0.85 V: strong CO signature; some $OCHCH_3$/OCCHO; $CO_3$ observed once.
  - Surface evidence at −0.76 V: CO increases, OCCO observed, OCCHO/OCCHOH present, HCOO still present.
  - FE at −0.90 to −0.80 V: formate and $H_2$ contributions rise (especially at −0.80), CO FE increases at intermediate potentials, while $C_2H_4$ and EtOH fall relative to most negative points.
  - Mechanistic alignment: Surface at −0.76 V already shows CO and OCCO emergence, which is consistent with the potential window where C2 formation is accessible; but FE at the nearby −0.80 to −0.90 shows increasing formate and $H_2$ and decreased $C_2H_4$ relative to −1.00 V. This could reflect:

- (i) small potential offsets (CO/OCCO may already appear at −0.76 but C2 product formation is sensitive to modest potential shifts), or
- (ii) differences in experimental conditions (electrolyte, current density, catalyst state) that change product outcomes despite similar surface motifs.
- Caveat: the spectroscopic presence of OCCO/CO* does not guarantee high C2 FE at the FE potential if subsequent hydrogenation/deoxygenation steps or desorption are rate-limiting.

- Surface −0.56 V and −0.36 V (compare to FE −0.80 to −0.90 V or treat as surface-only)
- Surface evidence: HCOO/HCOOH abundant at −0.36 and still present at −0.56; OCCHO and oxygenated C2 motifs also appear even at −0.36 V.
- FE at −0.80 (most positive FE point shown): formate FE increases and is a dominant contributor, H2 and CO large.
- Mechanistic alignment: Independent surface detection of HCOO/HCOOH at mild potentials is qualitatively consistent with the FE trend that formate becomes a major product at less negative potentials (FE shows increasing formate as potential becomes less cathodic). This supports the assignment that a formate pathway is accessible at mild overpotentials.
- Caveat: the FE figure's least negative potential is still around −0.80 V (more negative than −0.36 and −0.56), so direct quantitative comparison is only contextual.

B. Comparison table (condensed)

Potential | Surface-state evidence (inferred motifs; tentative) | Major FE products (figure) | Mechanistic interpretation | Agreement / discrepancy | Key limitations
---------------------------------------------------------------------------------------------------------------------------------------------------------
−1.12 (surface) ↔ −1.10 (FE)
| CO* frequent; OCCO (CO–CO); OCCHO/OCCHOH; CH/CHO; some complex O-C2
| Large C2H4; CH4 and H2 substantial; EtOH measurable
| CO* + OCCO consistent with CO-dimerization to C2; OCCHO/OH motifs compatible with ethanol pathway; CH/CHO consistent with CH4
| Good qualitative agreement: spectroscopic evidence for CO-dimerization and C1 hydrogenation matches C2H4 & CH4 FE. | OCCO inferred in few runs; species abundance not quantified; experimental conditions may differ.
---------------------------------------------------------------------------------------------------------------------------------------------------------
−0.95 (both)
| CO* dominant; OCHCH3/OCHCH2 present; HCOO minor; CH occasional
| C2H4 major; EtOH present; CH4 & H2 moderate
| CO* and oxygenated C2 motifs support concurrent pathways: CO-dimer → C2 and hydrogenation to oxygenates (EtOH); CH indicates C1 hydrogenation branch.
| Qualitatively consistent: both C2 hydrocarbons and oxygenates observed in FE, matching mixed surface motifs. | O-C2 labels ambiguous; detection ≠ quantification; transient vs on-path unclear.
---------------------------------------------------------------------------------------------------------------------------------------------------------
−0.85 / −0.76 (surface) ↔ −0.90/−0.80 (FE)
| At −0.76: CO rises, OCCO seen, OCCHO/OCCHOH; HCOO still present. At −0.85: strong CO, some OCHCH3.
| FE near −0.90 to −0.80: formate & H2 increase; C2H4 declines vs most negative potentials; CO FE non-negligible
| Surface shows onset of CO-dominated chemistry and OCCO (C2 precursors) at these potentials; FE shows a shift toward formate/H2 at the less negative FE points.
| Partial agreement: CO/OCCO appearance supports possible C2 formation, but FE shows C2 reduction is not dominant at all nearby potentials. Discrepancy could be due to potential offsets or kinetics. | Potentials not identical; spectral intermittency; reaction kinetics and site heterogeneity can decouple detected intermediates from product distribution.
---------------------------------------------------------------------------------------------------------------------------------------------------------

−0.56 / −0.36 (surface only) ↔ FE less negative region context
| HCOO/HCOOH prominent; oxygenated C2 motifs sometimes present; little CO
| FE at the least negative measured point (−0.80) shows large formate and H2 contributions
| Surface detection of formate at mild potentials supports formate pathway dominance at low overpotential. | Consistent qualitatively: formate pathway signature on surface matches higher formate FE at less negative potentials in FE trend. | FE potential is more negative than these surface points; direct match lacking; conditions could differ.

C. Concise potential-dependent pathway interpretation (supported vs plausible vs speculative)

- Supported by combined evidence (surface motifs + FE trends)
1) At more negative potentials (≈ −1.0 to −1.1 V) the surface is CO-rich (CO*) and OCCO-type motifs appear; FE shows high C2H4 and measurable EtOH and CH4. This supports CO-mediated C–C coupling (CO* → OCCO* → hydrogenated C2 intermediates) as a key route to C2 products; simultaneous CH/CHO signals and CH4 FE support competition with C1 hydrogenation.
2) At milder potentials the formate motif (HCOO*/HCOOH) is spectroscopically strong and FE shows increasing formate — consistent with a parallel formate route dominating at lower overpotentials.

- Plausible (consistent but not uniquely proven)
1) OCCHO/OCCHOH and OCHCH3/OCHCH2 motifs are hydrogenated CO–CO derivatives and can be branching intermediates to either ethanol (if O is retained and further hydrogenated) or ethylene (if deoxygenation/C–O scission occurs). The FE pattern (greater ethylene than ethanol at most negative potentials) suggests that for many surface motifs deoxygenation pathways compete effectively.
2) The intermittency of OCCO detection but steady high C2 FE suggests OCCO may be transient (short-lived) and low in steady-state spectroscopic signal yet kinetically important.

- Speculative (require further validation)
1) Assigning OHCCOH or specific enediol/diol structures as committed ethanol precursors is speculative given spectral ambiguity and single detections.
2) Any precise sequence (which protonation site on OCCO controls ethylene vs ethanol branching) is speculative without isotope/time-resolved data and DFT vibrational assignments.

D. Why detecting an intermediate does not equal high FE or high flux
- Detection = presence (or a spectroscopically visible population) at steady state; it does not measure turnover frequency, branching fractions, or whether the species is on-path or a spectator/poison.
- A species can be (i) spectator or strongly bound (trapping carbon without rapid conversion), (ii) low in population but high flux (short residence time) and thus spectroscopically weak, or (iii) localized to a minor fraction of surface sites that control selectivity but contribute little to bulk spectral signal.
- Thus CO* being spectroscopically prominent is consistent with high CO coverage but does not alone quantify how much of that CO* dimerizes versus desorbs or hydrogenates to CH4.

E. Consistency and main discrepancies (with possible causes)
- Agreements:
  - Rising CO*/OCCO signatures at more negative potentials align with increased C2H4 and C2 products in FE — supports a CO-dimerization channel to C2.
  - Formate/HCOO* presence at mild potentials aligns with higher formate FE at less negative potentials.
  - CH/CHO surface indicators at most negative potentials align with measurable CH4 FE.
- Discrepancies / unresolved points:
  - OCCO is inferred only intermittently while FE shows strong, reproducible C2 production — possible causes: OCCO is transient and spectroscopically weak; surface heterogeneity (only some domains produce C2 but dominate FE); different experimental conditions between spectroscopy runs and product measurement.
  - At potentials where surface still shows HCOO* (e.g., −0.76) FE may already show declining C2 or rising formate depending on the FE potential used for comparison — possible causes: small potential offsets, different electrolyte/current density, or kinetics making formate production competitive even when CO* is present.

F. New mechanistic insights from the comparison and testable hypotheses

1) Insight: CO* coverage appears necessary but not sufficient for high ethylene FE.

- Hypothesis: High steady-state CO* density enables OCCO formation, but subsequent competition between hydrogenation (→ oxygenated C2) and C–O scission (→ ethylene) determines C2 product distribution; the rate of deoxygenation relative to hydrogenation is the selectivity gate.

- Test: Time-resolved operando spectroscopy during potential steps with simultaneous product analysis; DFT/kinetic modeling of hydrogenation vs C–O scission barriers for representative OCCHO/OCHCHx species.

2) Insight: Formate and CO pathways coexist over a broad potential range; selectivity shifts reflect kinetic competition rather than a discrete mechanistic switch.

- Hypothesis: Surface heterogeneity (oxide-derived sites vs metallic Cu facets) spatially partitions activity (some domains favor HCOO*, others CO*), producing mixed FE and mixed spectral motifs.

- Test: Spatially resolved operando spectroscopy (micro-Raman/SERS mapping) and post-reaction surface characterization (XPS/XAS) to correlate domain composition with product formation; isotopically labelled CO2 to verify carbon incorporation into distinct surface motifs.

3) Insight: Intermittent detection of OCCO and O-rich C2 motifs with robust C2 FE suggests key C–C coupling intermediates are short-lived or low-population, while hydrogenated oxygenates may be longer-lived spectroscopic sinks.

- Hypothesis: OCCO* rapidly converts (low steady-state concentration) to OCCHO*/OCHCHx which are spectroscopically easier to detect; the latter's relative stability biases product distribution toward oxygenates unless C–O scission is facile.

- Test: Isotopic labeling (13CO/13CO2) with transient SERS/SEIRAS to follow C–C bond formation kinetics; correlate intermediate lifetimes with evolving product spectra.

G. The three most important mechanistic insights (each with observation, interpretation, alternative, evidence needed)

1) Insight 1

- Observation: At −1.12 (surface) CO* and OCCO are inferred; FE at −1.10 shows strong C2H4 and measurable EtOH and CH4.
- Mechanistic interpretation: CO-mediated C–C coupling (CO* → OCCO*) is operative at most negative potentials and is linked to the observed high C2H4 FE; OCCHO/OCHCHx motifs are plausible hydrogenated intermediates that can diverge to ethylene or ethanol.
- Alternative explanation: OCCO/OCCHO signals might reflect a minor domain or transient species while most C2H4 production proceeds via a different surface ensemble or catalytic site not captured strongly in the spectra.
- Evidence needed: Time-resolved correlation between OCCO/OCCHO appearance and C2 product formation (operando spectroscopy + fast product detection), and DFT vibrational assignments of candidate OCCO/OCCHO structures.

2) Insight 2

- Observation: HCOO/HCOOH motifs are prominent at mild potentials (−0.36, −0.56) and FE at less negative potentials shows increased formate FE.
- Mechanistic interpretation: A parallel formate pathway competes with the CO pathway across potentials; at milder overpotentials the formate branch dominates product distribution.
- Alternative explanation: Surface HCOO* detections could be from product readsorption/accumulation rather than a dominant on-path intermediate.
- Evidence needed: 13CO2 tracing to verify carbon in detected surface HCOO* derives from CO2 and potential-step kinetics linking HCOO* dynamics to formate production rates.

3) Insight 3

- Observation: O-containing C2 motifs (OCHCH3/OCHCH2, OCCHO variants) appear across potentials, sometimes when FE shows ethylene >> ethanol.
- Mechanistic interpretation: Detection of oxygenated C2 motifs indicates hydrogenated CO–CO intermediates are accessible; selectivity toward ethylene versus ethanol likely depends on the ease of C–O scission versus further hydrogenation (i.e., branching occurs at hydrogenated acyl/enolate species).
- Alternative explanation: Those O-rich motifs may be spectroscopically stable off-path species (spectator or re-adsorbed products) and not direct precursors to ethylene/ethanol.
- Evidence needed: Isotope labeling (H/D and 13C) coupled to operando vibrational spectroscopy and product analysis to track whether these motifs convert into gas-phase ethylene/ethanol and to measure their lifetimes.

H. 150–200 word Results/Discussion paragraph (suitable for a paper)

At increasingly cathodic potentials the oxide-derived Cu surface transitions from a formate-rich state to one dominated by adsorbed CO and transient CO–CO coupled motifs, and this shift correlates with the measured product distribution. Operando spectral reconstructions infer HCOO*/HCOOH at mild overpotentials and rising CO* and OCCO-like species near −1.0 V, while FE measurements show formate and H2 favored at less negative potentials and pronounced C2H4 (with accompanying EtOH and CH4) at the most negative potentials. The combined data support a model in which CO coverage is a gating variable for C–C coupling: sufficient CO* density enables OCCO formation and downstream hydrogenated OCCHO/OCHCHx intermediates, which can either lose oxygen (favoring ethylene) or be further hydrogenated (favoring ethanol). Importantly, spectroscopic visibility does not equate to kinetic dominance: OCCO may be transient or low in steady-state concentration yet control C2 flux, whereas more stable oxygenated C2 species may accumulate without guaranteeing desorption as ethanol. Discriminating these branching steps will require time-resolved isotope labeling and correlated operando product detection.

I close with methodological caveats: the inferred species are tentative spectral assignments, spectroscopic frequency of detection ≠ surface coverage or turnover, and FE data were taken under possibly different experimental conditions — these limit quantitative validation but not the qualitative mechanistic inferences above.